\documentstyle[tighten,preprint,eqsecnum,aps,prd]{revtex}
\begin{document}
\draft

\hyphenation{
mani-fold
mani-folds
}


\def\Bbb{\bf}

\def\BbbR{{\Bbb R}}
\def\BbbZ{{\Bbb Z}}

\def\half{{1\over2}}
\def\casehalf{{\case{1}{2}}}

\def\mink{{M^{2+1}}}
\def\KB{{\hbox{\it KB}}}
\def\bbase{{\cal B}}

\def\su{{\hbox{${\rm SU}(1,1)$}}}

\def\io{{\hbox{${\rm IO}(2,1)$}}}
\def\ioc{{\hbox{${\rm IO}_0(2,1)$}}}
\def\iop{{\hbox{${\rm IO}_{\rm P}(2,1)$}}}
\def\iot{{\hbox{${\rm IO}_{\rm T}(2,1)$}}}
\def\iotp{{\hbox{${\rm IO}_{\rm TP}(2,1)$}}}

\def\o{{\hbox{${\rm O}(2,1)$}}}
\def\oc{{\hbox{${\rm O}_0(2,1)$}}}
\def\op{{\hbox{${\rm O}_{\rm P}(2,1)$}}}
\def\ot{{\hbox{${\rm O}_{\rm T}(2,1)$}}}
\def\otp{{\hbox{${\rm O}_{\rm TP}(2,1)$}}}

\def\ba{{\cal A}}

\def\miso{{{\cal M}}}
\def\moo{{\miso_{0,0}}}
\def\mto{{\miso_{{\rm T},0}}}
\def\mpo{{\miso_{{\rm P},0}}}
\def\mttwo{{\miso_{{\rm T},{\rm TP}}}}
\def\mptwo{{\miso_{{\rm P},{\rm TP}}}}

\def\misotorus{{\miso^{\rm torus}}}
\def\misotorusconn{{\miso^{\rm torus}_{0,0}}}
\def\misotorusone{{\miso^{\rm torus}_{0,{\rm TP}}}}
\def\misotorustwo{{\miso^{\rm torus}_{{\rm TP},{\rm TP}}}}

\def\arsinh{{\mathop{\rm arsinh}\nolimits}}


\def\mathpounds{{\mathchoice{{\hbox{\pounds}}}{{\hbox{\pounds}}}%
{{\hbox{\scriptsize\pounds}}}{{\hbox{\scriptsize\pounds}}}}}


\preprint{\vbox{\baselineskip=12pt
\rightline{WISC--MILW--95--TH--15}
\rightline{gr-qc/9505026}}}
\title{Witten's 2+1 gravity on
${\BbbR} \times\hbox{(Klein bottle)}$}
\author{Jorma Louko\cite{jorma}}
\address{
Department of Physics,
University of
Wisconsin--Milwaukee,
\\
P.O.\ Box 413,
Milwaukee, Wisconsin 53201, USA}
\date{May 1995}
\maketitle
\begin{abstract}%
Witten's formulation of 2+1 gravity is investigated on the
nonorientable three-manifold $\BbbR \times\hbox{(Klein bottle)}$.
The gauge group is taken to consist of all four components of the
2+1 Poincare group. We analyze in detail several components
of the classical solution space, and we show that from four
of the components one can recover
nondegenerate spacetime metrics.
In particular, from one component we recover metrics
for which the Klein bottles are spacelike.
An action principle is formulated for bundles satisfying
a certain orientation compatibility property, and the
corresponding components of the classical solution
space are promoted into a phase space. Avenues
towards quantization are briefly discussed.
\end{abstract}%

\widetext

\section{Introduction}
\label{sec:intro}

The observation that vacuum Einstein gravity in 2+1 spacetime
dimensions has no local dynamical degrees of
freedom\cite{deser1,deser2} has created interest in 2+1 gravity as an
arena where quantum gravity can be investigated without many of the
technical complications that are present in 3+1 spacetime dimensions.
Several formulations of 2+1 dimensional gravity have been given, both
classically and quantum mechanically, and it has become an issue to
understand what the differences in these formulations are. For
reviews, see Refs.\cite{carlip-water,carlip-lect}.

In this paper we shall consider Witten's connection formulation of
2+1 gravity\cite{achu,witten1}. For spacetimes with the topology
$\BbbR\times \Sigma$, where $\Sigma$ is a closed oriented surface,
Witten's formulation and its correspondence with the metric theory
have been extensively
discussed\cite{witten1,moncrief1,moncrief2,mess,five-a,
AAbook2,carlip1,amano,anderson,marolf,louma};
a comprehensive list of references can be found in
Refs.\cite{carlip-water,carlip-lect}. The purpose of the
present paper is to extend this discussion to the
nonorientable spacetime $\BbbR\times \KB$, where $\KB$ stands for the
Klein bottle.

For a closed oriented surface $\Sigma$, a connection formulation
that reproduces the standard metric solutions on $\BbbR\times \Sigma$
can be defined as the theory of flat connections on principal $\ioc$
bundles over $\BbbR\times\Sigma$, where $\ioc$ is the component of
the identity of the 2+1 dimensional
Poincare group~$\io$\cite{witten1,mess}.
For $\BbbR\times \KB$ a similar connection theory with
the gauge group $\ioc$ is well-defined,
as we shall see, but it is not clear whether
this theory in any sense describes nondegenerate spacetime
metrics on $\BbbR\times\KB$. This motivates us to study
the theory in which the gauge group consists of
all of $\io$. It will be shown that nondegenerate,
flat metrics on $\BbbR\times\KB$ can be recovered from four connected
components of the classical solution space of this theory, and from
one of the components we recover metrics that admit
spacelike Klein bottles. We shall also give an action principle for
certain components of the theory, including these four. The action is
sufficiently general to accommodate both the nonorientability of the
manifold and the fact that the bundles are nontrivial in a manner
involving the disconnected components of $\io$.

We begin in Section \ref{sec:connection} by describing the connection
theory. This theory is a modest generalization of
that given in Ref.\cite{witten1},
in that the three-manifold $M$ may be nonorientable
and the gauge group is the full 2+1 Poincare group $\io$.
We first discuss the kinematics of connections on $\io$ bundles
over~$M$, devoting special attention to gauge
transformations and to the recovery of a spacetime metric on~$M$. The
equations of motion are then introduced by the requirement
that the connection be flat.
We next define a bundle to be orientation
compatible iff the (potential) nonorientability of the base space
intertwines with the (potential) nontriviality of the bundle in a
certain way, and we construct for such bundles an action principle
from which the equations of motion can be derived.
$\ioc$ bundles over oriented manifolds are orientation compatible,
and for them our action reduces to that given in
Refs.\cite{witten1,amano}.

In Section \ref{sec:kbconnsol} we specialize to $M=\BbbR\times \KB$,
and we analyze in detail several of the connected components of the
classical solution space. In Section \ref{sec:metrics} we demonstrate
that four of these components contain points from which one recovers
nondegenerate metrics on~$M$. The corresponding bundles are
nontrivial and orientation compatible. In particular, from one
component we recover metrics for which the induced metric on $\KB$ is
positive definite.

Section \ref{sec:quantization} offers some remarks on quantization.
For the orientation compatible bundles, the action principle of
Section \ref{sec:connection} endows the classical solution space,
after excision of certain singular subsets,
with a symplectic structure.
This enables us to interpret these components of the solution space
as cotangent bundles (again, after excision of certain singular
subsets) and to apply methods of geometric
quantization\cite{woodhouse,AAbook-geom}. The results are
qualitatively similar to those found for the $\ioc$ connection theory
on the manifold $\BbbR\times T^2$ in Ref.\cite{louma}.

Section \ref{sec:discussion} contains a brief summary
and discussion. Some notation and
facts about $\io$ and the Klein bottle have been collected into
appendices \ref{app:io} and~\ref{app:kb}, and certain technical
details are postponed to Appendix~\ref{app:mmpo}.
In Appendix \ref{app:torustheory} we discuss briefly
the $\io$ connection theory on the manifold $\BbbR \times T^2$.

\section{2+1 gravity in the $\io$ connection formulation}
\label{sec:connection}

In this section we describe the theory of flat connections on
principal $\io$ bundles over a three-manifold~$M$, and its
correspondence to 2+1 dimensional Einstein gravity.
The notation and some facts about $\io$ have been
collected into Appendix~\ref{app:io}.

\subsection{Kinematics}
\label{subsec:kinematics}

The kinematical arena of the theory consists of a (connected,
paracompact, Hausdorff, $C^\infty$) differentiable
three-manifold~$M$, a principal $\io$ bundle $P$ over~$M$, and a
connection $\ba$ in~$P$. In a local chart
$(U_\alpha,\varphi_\alpha)$, the representative of $\ba$ is a
one-form ${\vphantom{\ba}}^\alpha\!\!\ba_a$ on $U_\alpha$ taking
values in the Lie algebra of $\io$. The lowercase Latin index is
understood as an abstract tensor index on~$M$. In terms of the basis
$\left\{ J_I, P_I \right\}$ of the Lie algebra of $\io$
(see Appendix~\ref{app:io}),
${\vphantom{\ba}}^\alpha\!\!\ba_a$ can be expanded as
\begin{equation}
{}^\alpha\!\!\ba_a =
{}^\alpha \! e_a^I P_I + {}^\alpha \!\! A_a^I J_I
\ \ ,
\label{baexpand}
\end{equation}
where ${}^\alpha \! e_a^I$ and ${}^\alpha \!\! A_a^I$ are
one-forms on~$U_\alpha$.
The local representative of the curvature two-form of $\ba$ is
\begin{equation}
{}^\alpha \! {\cal F}_{ab} =
2 \, {}^\alpha {\cal D} {}^{\phantom{I}}_{[a}
{}^\alpha \! e_{b]}^I P_I
+ {}^\alpha \!\! F^I_{ab} J_I
\ \ ,
\label{iocurv}
\end{equation}
where
\begin{equation}
{}^\alpha \! F^I_{ab} =
2 \partial^{\phantom{I}}_{[a} {}^\alpha \!\! A^I_{b]} +
\epsilon^I{}_{JK} \, {}^\alpha \!\! A^J_a \,
{}^\alpha \!\! A^K_b
\ \ ,
\label{F}
\end{equation}
and ${}^\alpha {\cal D}_a$ is the
$\ba$-dependent derivative operator in $U_\alpha$ defined by
\begin{equation}
{}^\alpha {\cal D}_a v^I =
\partial_a v^I + \epsilon^I{}_{JK} \,
{}^\alpha \!\! A^J_a v^K
\ \ .
\label{cald}
\end{equation}
Note that ${}^\alpha {\cal D}_a$ acts on the
Lorentz indices but not on the tensor indices on~$M$.

If $(U_\alpha,\varphi_\alpha)$ and $(U_\beta,\varphi_\beta)$ are two
local charts with nonempty $U_\alpha\cap U_\beta$, the
representatives of $\ba$ on $U_\alpha\cap U_\beta$ are related by the
gauge transformation\cite{koba}
\begin{equation}
{\vphantom{\ba}}^\beta\!\!\ba_a =
{\rm ad} (\psi_{\alpha\beta}^{-1}) \,
{\vphantom{\ba}}^\alpha\!\!\ba_a
+ \psi_{\alpha\beta}^{-1} d\psi_{\alpha\beta}
\ \ ,
\label{gaugetrans}
\end{equation}
where $\psi_{\alpha\beta} = \varphi_\alpha \varphi_\beta^{-1}$ is
the transition function.
Writing $\psi_{\alpha\beta} = (R_{\alpha\beta}, w_{\alpha\beta})$
in the notation of Appendix~\ref{app:io},
(\ref{gaugetrans}) gives the transformations
\begin{mathletters}
\label{gaugetransae}
\begin{eqnarray}
{}^\beta \!\! A^I_a K_I &=&
R_{\alpha\beta}^{-1}
\left( {}^\alpha \!\! A^I_a K_I \right)
R_{\alpha\beta}
+ R_{\alpha\beta}^{-1} \partial_a R_{\alpha\beta}
\nonumber
\\
&=&
\det\left(R_{\alpha\beta}\right)
{\left(R^{-1}_{\alpha\beta}\right)}^I
{\vphantom{\left(R^{-1}_{\alpha\beta}\right)}}_J \,
{}^\alpha \!\! A^J_a K_I
+ R_{\alpha\beta}^{-1} \partial_a R_{\alpha\beta}
\ \ ,
\label{gaugetransa}
\\
{}^\beta \! e^I_a  &=&
{\left( R_{\alpha\beta}^{-1}\right)}^I
{\vphantom{\left(R_{\alpha\beta}^{-1}\right)}}_J
\left(
{}^\alpha \! e^J_a
+ {}^\alpha {\cal D}_a w_{\alpha\beta}^J
\right)
\ \ .
\label{gaugetranse}
\end{eqnarray}
\end{mathletters}%
It follows that
\begin{equation}
{}^\beta \! F^I_{ab} =
\det\left(R_{\alpha\beta}\right)
{\left(R^{-1}_{\alpha\beta}\right)}^I
{\vphantom{\left(R^{-1}_{\alpha\beta}\right)}}_J
\, {}^\alpha \! F^J_{ab}
\ \ .
\label{Ftrans}
\end{equation}

${}^\alpha \! e^I_a$ is interpreted as a triad on~$U_\alpha$,
and the corresponding metric on $U_\alpha$ is
${}^\alpha \! g_{ab} = \eta_{IJ} \, {}^\alpha \! e^I_a \,
{}^\alpha \! e^I_b$. If
${}^\alpha \! e^I_a$ is nondegenerate, ${}^\alpha \! g_{ab}$ is
nondegenerate with signature $(-++)$. The collection of metrics
$\left\{ {}^\alpha \! g_{ab} \right\}$ does not
in itself define a
metric on~$M$, since in the overlaps $U_\alpha\cap U_\beta$ the
metrics ${}^\alpha \! g_{ab}$ and ${}^\beta \! g_{ab}$ need
not coincide. To proceed, we use the fact
that the bundle $P$ is reducible to an $\o$ bundle
over~$M$.\footnote{This follows from Refs.\cite{husemoller,massey2}
and the observation that the coset space $\io/\o$ is homeomorphic
to~$\BbbR^3$. I~thank Domenico Giulini for pointing this out.}
This means that there exists an open cover of $M$ and an associated
system of local charts such that the transition functions
take values in $\o$; conversely, any such system of local
charts defines an $\o$ reduction of $P$\cite{koba}. Now, in a
system of local charts corresponding to an $\o$ reduction
of~$P$, (\ref{gaugetranse}) shows that the metrics
${}^\alpha \! g_{ab}$ coincide in the overlaps, and the collection
$\left\{ {}^\alpha \! g_{ab} \right\}$
therefore defines a (possibly degenerate) metric $g_{ab}$
on~$M$. If ${}^\alpha \! e^I_a$ are nondegenerate
for every~$\alpha$,
$g_{ab}$ is nondegenerate with signature $(-++)$. The quantities
${}^\alpha \!\! A^I_aJ_I$ can be interpreted as the local
representatives of a
connection in the reduced $\o$ bundle, and the local representatives
of the curvature of this $\o$ connection are
${}^\alpha \! F^I_{bc}J_I$. Similarly, ${}^\alpha \! e^I_a$ can be
interpreted as triad fields associated with this reduced bundle.

The metric $g_{ab}$ obtained in this fashion depends on the
$\o$ reduction of~$P$. To investigate this dependence, we
shall for the remainder of this subsection denote by
$\left\{(U_\alpha, \varphi_\alpha) \right\}$ and $\left\{(U_\alpha,
{\bar \varphi}_\alpha) \right\}$ two systems of local charts
of $P$ corresponding to two $\o$ reductions of~$P$.
The transition
functions $\psi_{\alpha\beta} = \varphi_\alpha \varphi_\beta^{-1}$
and ${\bar \psi}_{\alpha\beta} = {\bar \varphi}_\alpha {\bar
\varphi}_\beta^{-1}$ thus take values in $\o$. Let $\theta_\alpha =
\varphi_\alpha {\bar \varphi}^{-1}_\alpha$ be the transition function
between the charts $(U_\alpha, {\bar \varphi}_\alpha)$ and
$(U_\alpha, \varphi_\alpha)$. In a nonempty $U_\alpha \cap U_\beta$,
the consistency of the transition functions implies
\begin{equation}
{\bar \psi}_{\alpha\beta} =
\theta^{-1}_\alpha \psi_{\alpha\beta} \theta_\beta
\ \ .
\label{psitrans}
\end{equation}
In the notation of Appendix~\ref{app:io}, we write
$\psi_{\alpha\beta} = \left( S_{\alpha\beta}, 0 \right)$,
$\theta_\alpha = \left(R_\alpha, w_\alpha \right)$.
As ${\bar \psi}_{\alpha\beta}$ takes values in $\o$,
(\ref{psitrans}) implies
\begin{equation}
w_\beta = S^{-1}_{\alpha\beta} w_\alpha
\ \ .
\label{wtrans}
\end{equation}
Let now ${}^\alpha \! e^I_a$ and ${}^\alpha \! {\bar e}^I_a$ be
the triads associated respectively with the charts $(U_\alpha,
\varphi_\alpha)$ and $(U_\alpha, {\bar \varphi}_\alpha)$, and let us
assume that ${}^\alpha \! e^I_a$ are nondegenerate
for every~$\alpha$. Then there
exists in each $U_\alpha$ a unique vector field ${}^\alpha \! f^a$
such that
\begin{equation}
w_\alpha^I = {}^\alpha \! e^I_a \, {}^\alpha \! f^a
\ \ .
\label{fdef}
\end{equation}
In a nonempty $U_\alpha\cap U_\beta$, equations (\ref{gaugetranse}),
(\ref{wtrans}), and (\ref{fdef}) show that ${}^\alpha \! f^a$ and
${}^\beta \! f^a$ coincide, and the collection
$\left\{ {}^\alpha \! f^a \right\}$ thus defines a vector field
$f^a$ on~$M$. From
(\ref{gaugetranse}) one then obtains
\begin{equation}
{}^\alpha \! {\bar e}^I_a =
{\left( R_\alpha^{-1}\right)}^I
{\vphantom{\left(R_\alpha^{-1}\right)}}_J
\left(
{}^\alpha \! e^J_a
+ \mathpounds_f {}^\alpha \! e^J_a
+ f^b \epsilon^J{}_{KL} \,
{}^\alpha \!\! A^K_b \, {}^\alpha \! e^L_a
+ 2 f^b \, {}^\alpha {\cal D} {}^{\phantom{I}}_{[a}
{}^\alpha \! e_{b]}^J
\right)
\ \ ,
\label{transebar}
\end{equation}
where $\mathpounds_f$ is the Lie derivative with respect to $f^a$.
For the metrics ${}^\alpha \! {\bar g}_{ab} = \eta_{IJ} \, {}^\alpha
\! {\bar e}^I_a \, {}^\alpha \! {\bar e}^I_b$ and ${}^\alpha \!
g_{ab} = \eta_{IJ} \, {}^\alpha \! e^I_a \, {}^\alpha \! e^I_b$,
(\ref{transebar}) gives to first order in $f^a$ the relation
\begin{equation}
{}^\alpha \! {\bar g}_{ab} =
{}^\alpha \! g_{ab} + \mathpounds_f \, {}^\alpha \! g_{ab}
+ 2 f^c \left(
{}^\alpha \! e_{Ja} \,
{}^\alpha {\cal D} {}^{\phantom{I}}_{[b}
{}^\alpha \! e_{c]}^J
+
{}^\alpha \! e_{Jb} \,
{}^\alpha {\cal D} {}^{\phantom{I}}_{[a}
{}^\alpha \! e_{c]}^J
\right)
+ O\left( {(f)}^2 \right)
\ \ .
\label{deltag}
\end{equation}

\subsection{Dynamics}
\label{subsec:dynamics}

The dynamics of the theory consists of the statement that the
connection $\ba$ is flat. In a local chart this amounts by
(\ref{iocurv}) to the equations of motion
\begin{mathletters}
\label{eom}
\begin{eqnarray}
{}^\alpha \! F^I_{ab} &=& 0
\ \ ,
\label{eom1}
\\
{}^\alpha {\cal D} {}^{\phantom{I}}_{[a}
{}^\alpha \! e_{b]}^I &=& 0
\ \ .
\label{eom2}
\end{eqnarray}
\end{mathletters}%
In a system of local charts corresponding to an $\o$
reduction of~$P$,  (\ref{eom1}) says that the connection on the
reduced bundle is flat, and (\ref{eom2}) can then be understood as a
compatibility condition for the $\o$ connection and the triad.

In the theory where $\io$ is replaced by $\ioc$,
an $\oc$ reduction of $P$ defines on $M$ the
three-form $e^{\phantom{I}}_{I[a} F^I_{bc]}$.
If $M$ is orientable, integrating this three-form over $M$ yields
then an action functional from which the equations of motion
(\ref{eom}) can be derived\cite{witten1,AAbook2,amano,romano}. We
shall now show that this action can be generalized
to a class of bundles within our $\io$ theory, including
certain bundles for nonorientable~$M$.
The idea is to define on $M$ a
density analogous to $e^{\phantom{I}}_{I[a} F^I_{bc]}$ by taking
advantage of the factor $\det\left(R_{\alpha\beta}\right)$
in~(\ref{Ftrans}).

Let $P$ be a principal $\io$ bundle over~$M$. Let $P'$ be the
principal ${\BbbZ}_2$ bundle over $M$ that is obtained by collapsing
the fibers in $P$ into ${\BbbZ}_2$ with the homomorphism
$\io\to{\BbbZ}_2; (R,w)\mapsto\det(R)$. Let $x\in M$, and let
$\pi_1(M,x)$ be the homotopy group of $M$ with the base point~$x$. As
${\BbbZ}_2$ is discrete and Abelian, lifting closed paths in $M$ with
the base point $x$ into paths in $P'$ defines the holonomy map
$h_x\colon \pi_1(M,x)\to{\BbbZ}_2$\cite{koba}. By construction $h_x$
is a group homomorphism. We say that $P'$ is {\it orientation
compatible\/} iff for every $x\in M$, $h_x$ takes the orientation
preserving homotopy classes in $\pi_1(M,x)$ to $1$ and the
orientation reversing homotopy classes (if any) to~$-1$. It is clear
that an equivalent definition is to require $h_x$ to have this
property for some $x\in M$. We say that $P$ is orientation compatible
iff $P'$ is orientation compatible.

For the rest of this subsection we assume $P$ to be orientation
compatible. As in subsection~\ref{subsec:kinematics}, we introduce a
system of local charts
$\left\{(U_\alpha,\varphi_\alpha)\right\}$ corresponding to an $\o$
reduction of~$P$. The index set in which $\alpha$ takes values is
denoted by~$I$. Without loss of generality we can assume this system
to be chosen such that there is an atlas
$\left\{(U_\alpha,\sigma_\alpha)\right\}$ of $M$ with the local
manifold charts $\sigma_\alpha \colon U_\alpha\to {\BbbR}^3$.

Let $P'$ be the principal ${\BbbZ}_2$ bundle over $M$ that is
obtained from $P$ by collapsing the fibers into ${\BbbZ}_2$ as above.
Let $\left\{(U_\alpha,\varphi'_\alpha)\right\}$ be the system of
local charts of $P'$ induced by the system
$\left\{(U_\alpha,\varphi_\alpha)\right\}$. We now single out one
value of $\alpha$, say $\alpha=0$, and we shall use the
chart $(U_0,\varphi'_0)$ and the orientation of $U_0$ induced
by the manifold chart $\sigma_0$ as a reference
in terms of which the action will be defined.
This choice is analogous to choosing the orientation
on $M$ in the theory considered in Refs.\cite{witten1,amano}.

For every $\alpha$, choose a point $x_\alpha\in U_\alpha$. Let
$p'_\alpha \in P'$ be the point in the fiber over $x_\alpha$ such
that $\varphi'_\alpha(p'_\alpha)=1$. Let $\gamma_\alpha$ be a path in
$M$ starting from $x_\alpha$ and ending at~$x_0$. We now define the
numbers $\epsilon_{\gamma_\alpha}$ and $\eta_{\gamma_\alpha}$ in
${\BbbZ}_2$ in the following way. Let ${\tilde\gamma}_\alpha$ be the
lift of $\gamma_\alpha$ to $P'$ that starts from~$p'_\alpha$. We set
$\epsilon_{\gamma_\alpha}=1$ if the end point of
${\tilde\gamma}_\alpha$ is $p'_0$, and $\epsilon_{\gamma_\alpha}=-1$
otherwise. We set $\eta_{\gamma_\alpha}=1$ if $\gamma_\alpha$ takes
the orientation of $U_\alpha$ induced by the
manifold chart $\sigma_\alpha$ to
the orientation of $U_0$ induced by the
manifold chart $\sigma_0$, and
$\eta_{\gamma_\alpha}=-1$ otherwise. The orientation compatibility of
$P'$ guarantees that for a fixed~$\alpha$, the product
$\epsilon_{\gamma_\alpha}\eta_{\gamma_\alpha}$ is independent of the
choice of the path $\gamma_\alpha$ and the choice of the point
$x_\alpha\in U_\alpha$. The product
$\epsilon_{\gamma_\alpha}\eta_{\gamma_\alpha}$ therefore defines a
function $E \colon I\to{\BbbZ}_2$. In particular one has $E(0)=1$.

Let now $\ba$ be a connection in~$P$. In each $U_\alpha$, use the
local chart $(U_\alpha,\varphi_\alpha)$ as in subsection
\ref{subsec:kinematics} to define the three-form
\begin{equation}
{}^\alpha \! {\cal E}_{abc} =
E(\alpha) \,
{}^\alpha \! e^{\phantom{I}}_{I[a}
\,
{}^\alpha \! F^I_{bc]}
\ \ .
\label{eF}
\end{equation}
Let ${}^\alpha \! {\cal E}_{\bar{a}\bar{b}\bar{c}}$ be the components
of ${}^\alpha \! {\cal E}_{abc}$ in the manifold chart
$(U_\alpha,\sigma_\alpha)$; the overlined lowercase Latin indices are
concrete indices in this chart. In a nonempty $U_\alpha\cap U_\beta$,
it now follows from (\ref{gaugetranse}) and (\ref{Ftrans}) that
${}^\alpha \! {\cal E}_{\bar{a}\bar{b}\bar{c}}$ and ${}^\beta \!
{\cal E}_{\bar{a}\bar{b}\bar{c}}$ are related by the absolute value
of the Jacobian of the manifold transition function $\sigma_\alpha
\sigma_\beta^{-1}$. This means that the collection $\left\{ {}^\alpha
\! {\cal E}_{\bar{a}\bar{b}\bar{c}} \right\}$ defines on $M$ a
density ${\tilde{\cal E}}_{abc}$\cite{botttu}. We can thus integrate
${\tilde{\cal E}}_{abc}$ over $M$ to obtain the action
\begin{equation}
S = \casehalf \int_M {\tilde{\cal E}}
\ \ .
\label{action}
\end{equation}
If $M$ is noncompact, suitable fall-off conditions may need to
imposed, and suitable boundary terms may need to be
added to~(\ref{action}).
Using the local representatives of~${\tilde{\cal E}}_{abc}$, it is
straightforward to verify that the variation of (\ref{action})
gives, under suitable boundary conditions in the noncompact case,
the equations of motion~(\ref{eom}). It
is also clear how this discussion generalizes if $M$ is
replaced by a manifold with a boundary.

Our definition of the action (\ref{action}) relied on the system
$\left\{(U_\alpha, \varphi_\alpha) \right\}$ of local charts
corresponding to an $\o$ reduction of~$P$. To see how the action
depends on the choice of this system, let $\left\{(U_\alpha, {\bar
\varphi}_\alpha) \right\}$ be another such system as in subsection
\ref{subsec:kinematics}. Let us first suppose that the transition
function ${\bar \varphi}'_0 {(\varphi'_0)}^{-1}$ is the identity. The
difference of ${\tilde{\cal E}}_{abc}$ and ${\tilde{\bar{\cal
E}}}_{abc}$ is then a total divergence
whose components in the manifold chart
$(U_\alpha,\sigma_\alpha)$ are $E(\alpha)
\partial^{\phantom{I}}_{[\bar{a}}
({}^\alpha \! F^I_{{\bar b}{\bar c}]} w_\alpha)$;
this divergence is a well-defined density by virtue of
(\ref{Ftrans}) and~(\ref{wtrans}). Given suitable fall-off/boundary
conditions, the action defined in terms of the new system of local
charts is therefore equal to the old action. Finally, if
${\bar \varphi}'_0 {(\varphi'_0)}^{-1}$ is not the identity, the two
actions differ only by an overall sign.

If $M$ is orientable, the construction of the action simplifies
considerably. In this case orientation compatibility is equivalent to
the triviality of~$P'$, and one can choose the local charts
of $P$ so as to induce a global chart of~$P'$. One can also
choose $\left\{(U_\alpha,\sigma_\alpha)\right\}$ to be an oriented
atlas. The factor $E(\alpha)$ in (\ref{eF}) is then equal to $1$ for
all $\alpha$ and can be omitted: ${\tilde{\cal E}}_{abc}$ can be
thought of just as a three-form on $M$\cite{witten1,amano}.

Finally, note that when $M = \BbbR \times \Sigma$, where $\Sigma$ is
a closed two-manifold, one can perform a Hamiltonian decomposition of
the action~(\ref{action}), in close analogy with the decomposition in
the $\ioc$ theory on orientable
manifolds\cite{witten1,AAbook2,romano}. The Hamiltonian form of the
action defines on the fields a symplectic structure, which can be
pulled back into a symplectic structure on (smooth subsets of) the
solution space. We shall discuss this explicitly in Section
\ref{sec:quantization} in the case where $\Sigma$ is the Klein
bottle.

\subsection{Solutions}
\label{subsec:connsol}

We shall now recall how to describe the solution space of the
connection theory in terms of the fundamental group of~$M$.

Let for the moment $G$ be a general Lie group, let $P$ be a principal
$G$ bundle over $M$, and let $\ba$ be a flat connection on~$P$.
Choose a point $x\in M$, and choose a point $p\in P$ in the fiber
over~$x$. As $\ba$ is flat, lifting closed paths in $M$ with the
base point $x$ into horizontal paths in $P$ starting at $p$
defines the holonomy
map~$\Phi_p \colon \pi_1(M,x)\to G$\cite{koba},
which is by construction a group homomorphism.
If $q\in P$ is another
point in the fiber over~$x$, one can write $q=pg$ for some $g\in G$,
and the holonomy maps are related by $\Phi_{pg}=g\Phi_p g^{-1}$.
Thus, $\ba$ defines a $G$ conjugacy class of elements in
$\hbox{Hom}(\pi_1(M,x),G)$, that is, a point in the quotient space
$\hbox{Hom}(\pi_1(M,x),G)/G$. Conversely, given
$\Phi\in\hbox{Hom}(\pi_1(M,x),G)$, there exists a principal $G$
bundle $P$ over $M$ and a flat connection $\ba$ on $P$ such that
$\Phi/G$ is the conjugacy class of homomorphisms defined by the
holonomy maps of~$\ba$, and the reconstruction of $P$ and $\ba$ from
$\Phi$ is unique up to isomorphisms\cite{barrett,lewa,jacobroma}.
Finally, given another point $x'\in M$, the holonomy maps
$\Phi'_{p'}\in\hbox{Hom}(\pi_1(M,x'),G)$ can be related to those in
$\hbox{Hom}(\pi_1(M,x),G)$ by introducing paths $\gamma$ in $M$ from
$x$ to~$x'$: the conjugacy classes are related just by
$\Phi'/G= (\Phi/G)\circ \Gamma_\gamma$, where
$\Gamma_\gamma \colon \pi_1(M,x') \to \pi_1(M,x)$
is the isomorphism induced by~$\gamma$.

We therefore have the following statement: {\it The space of flat
connections modulo bundle isomorphisms is parametrized by\/}
$\hbox{Hom}(\pi_1(M),G)/G$, {\it where\/} $\pi_1(M)$ {\it is the
(base point independent) fundamental group of\/}~$M$.

There are four issues that deserve a comment. Firstly, in the
present case of flat connections the
reconstruction\cite{barrett,lewa} of $P$ and $\ba$ from $\Phi$ takes
the following simple
form.\footnote{I thank Alan Rendall, Domenico Giulini, and Don Marolf
for pointing out this construction.}
Let ${\tilde M}$ be the universal covering space of~$M$,
and let $\pi_1(M)$ denote the base point independent
fundamental group of~$M$. There exists an action $\rho$ of
$\pi_1(M)$ on ${\tilde M}$ such that
the quotient space ${\tilde M}/\rho$ is homeomorphic
to~$M$\cite{massey}. We denote this action by ${\tilde x} \buildrel
\alpha \over \mapsto \rho_\alpha ({\tilde x})$, ${\tilde x}\in
{\tilde M}$, $\alpha\in\pi_1(M)$. Introduce the product bundle
${\tilde P}={\tilde M}\times G$, and let ${\tilde\ba}$ be the flat
connection on ${\tilde P}$ induced by the product structure. Given
$\Phi\in\hbox{Hom}(\pi_1(M),G)$, define the action ${\tilde\rho}$ of
$\pi_1(M)$ on ${\tilde P}$ by $({\tilde x}, g) \buildrel \alpha \over
\mapsto \left(\rho_\alpha ({\tilde x}),
{\left(\Phi(\alpha)\right)}^{-1} g \right)$ for $\alpha\in\pi_1(M)$.
It is straightforward to show that the quotient space ${\tilde
P}/{\tilde\rho}$ is a principal $G$ bundle $P$ over~$M$, and that
${\tilde\ba}$ induces a flat connection $\ba$ on~$P$. To study the
holonomy map of $\ba$ in~$P$, let $x\in M$, and let ${\tilde
x}\in{\tilde M}$ be a point in the $\rho$-equivalence class that
defines~$x$. The quotient construction $M={\tilde M}/\rho$ defines an
isomorphism $f_{\tilde x} \colon \pi_1(M,x)\to\pi_1(M)$. Let $p\in P$
be the ${\tilde\rho}$-equivalence class of $({\tilde x}, e)\in
{\tilde P}$, where $e$ is the identity element in~$G$. Clearly $p$ is
in the fiber over~$x$. The holonomy map
$\Phi_p\in\hbox{Hom}(\pi_1(M,x),G)$ of $\ba$ can then be verified to
be the composition $\Phi\circ f_{\tilde x}$.

Secondly, we are here adopting the viewpoint that connections are
identified only under those bundle isomorphisms whose projection to
the diffeomorphism group ${\rm Diff}(M)$ is in the component of the
identity. In physical terms, this means treating the
diffeomorphisms connected to the identity as ``gauge" but the
diffeomorphisms disconnected from the identity as symmetries.
If one wanted to identify connections also under the
bundle isomorphisms that project into the disconnected components
of~${\rm Diff}(M)$, one would need to take the quotient of
$\hbox{Hom}(\pi_1(M),G)/G$ with respect to the full automorphism
group of~$\pi_1(M)$. By construction, $\hbox{Hom}(\pi_1(M),G)/G$ is
already invariant under the inner automorphisms
of~$\pi_1(M)$\cite{goldman1,goldman2}.

Thirdly, $\hbox{Hom}(\pi_1(M),G)/G$ consists of points
arising from {\it all\/} $G$ bundles over $M$
that admit flat connections. For
a given~$M$, there may be several such bundles, in which case any
single bundle only yields a subset of $\hbox{Hom}(\pi_1(M),G)/G$.

Fourthly, on some bundles it is natural to introduce additional
structure that is not invariant under all bundle isomorphisms. To
consider the case of interest for us, suppose that $G$ is not
connected, and let $G_0$ be the component of the identity in~$G$.
Suppose now that one introduces some structure that is
not invariant under those bundle isomorphisms that permute the
components in the fibers. For example, one could introduce a base
point $*$ in~$M$, and fix one component in the fiber over $*$ to be
associated with~$G_0$. One can then require that solutions related by
a bundle isomorphism should be regarded as equivalent only if the
isomorphism leaves this structure invariant. The solution space is
then $\hbox{Hom}(\pi_1(M),G)/G_0$. The space
$\hbox{Hom}(\pi_1(M),G)/G$ is recovered as the quotient of
$\hbox{Hom}(\pi_1(M),G)/G_0$ with respect to conjugation by
$\pi_0(G)=G/G_0$. In physical language, adopting
$\hbox{Hom}(\pi_1(M),G)/G_0$ as the solution space means that the
gauge transformations that are disconnected from the identity are
treated as symmetries rather than as gauge.

The above discussion specializes readily to our version of Witten's
2+1 gravity. Choosing to treat both the large (that is, disconnected
from the identity) gauge transformations and the large
diffeomorphisms of $M$ as symmetries rather than as gauge, we see
that the solution space is $\hbox{Hom}(\pi_1(M),\io)/\ioc$, where
$\pi_1(M)$ is the (base point independent) fundamental group of~$M$.

\subsection{Spacetime metrics}
\label{subsec:metrics}

Given a solution $\ba$ to the connection theory on the bundle~$P$,
and given a local chart $(U_\alpha,\varphi_\alpha)$ such that
the triad ${}^\alpha \! e_{b}^I$ is nondegenerate, equations
(\ref{eom}) imply that the metric
${}^\alpha \! g_{ab} = \eta_{IJ} \,
{}^\alpha \! e^I_a \, {}^\alpha \! e^I_b$ on $U_\alpha$ is flat,
{\it i.e.,} satisfies the 2+1 dimensional vacuum Einstein
equations\cite{witten1,AAbook2,romano}. If there exists a system of
local charts corresponding to an
$\o$ reduction of $P$ such that
${}^\alpha \! e_{b}^I$ is nondegenerate for every~$\alpha$,
the collection
$\left\{{}^\alpha \! e_{b}^I\right\}$ then defines a flat
Lorentz-metric $g_{ab}$ on~$M$.

By (\ref{deltag}) and~(\ref{eom2}), an infinitesimal change in the
system of local charts changes $g_{ab}$ by a Lie derivative
with respect to a vector field on~$M$. Conversely,
it is seen from (\ref{fdef}) that for any vector
field $f^a$ on $M$ there exists an infinitesimal change in the $\o$
reduction such that the infinitesimal change in $g_{ab}$ is
just~$\mathpounds_f g_{ab}$. Thus, at the infinitesimal level, a
solution of the connection theory that
yields a spacetime metric on $M$ yields a whole diffeomorphism
equivalence class of such metrics.

In order to consider the effect of
finite changes in the system of local
charts on the metric, more precise assumptions are required.
Progress in this direction has been made in the case
$M={\BbbR}\times\Sigma$, where $\Sigma$ is a closed oriented
two-manifold, under the assumption that the induced metric on
$\Sigma$ is
spacelike\cite{witten1,moncrief1,moncrief2,mess,carlip1,amano}.
In particular, Ref.\cite{mess} gives
a set of assumptions that allows a
precise control of the diffeomorphism equivalence classes of
metrics when $(M,g_{ab})$ is assumed to be a
domain of dependence of~$\Sigma$.
If $\Sigma$ may be non-spacelike, the situation appears
more open. Some examples and discussion in the special case
$\Sigma=T^2$ can be found in Ref.\cite{louma}.

\section{Connection solutions on
${\BbbR} \times\hbox{(Klein bottle)}$}
\label{sec:kbconnsol}

In this section we shall analyze the solution space of the connection
theory on $\BbbR\times\KB$. Some facts about $\KB$ and its
fundamental group $\pi_1(\KB):=\pi$ are collected in
Appendix~\ref{app:kb}.

We choose to treat both the large gauge transformations and the large
diffeomorphisms of $M$ as symmetries, in the way explained in
subsection~\ref{subsec:connsol}. Since
$\pi_1(\BbbR\times\KB)\simeq\pi_1(\KB)=\pi$, the solution space
$\miso$ is
\begin{equation}
\miso = \hbox{Hom}(\pi,\io)/\ioc
\ \ .
\label{miso}
\end{equation}

To obtain a convenient description of $\hbox{Hom}(\pi,\io)$,
recall from Appendix \ref{app:kb} that $\pi$
is generated by a pair of elements $(a,b)$ with the single
relation~(\ref{pirelation}).
A~point in $\hbox{Hom}(\pi,\io)$ is therefore uniquely specified
by the images of the two generators, and the images must satisfy the
relation arising from~(\ref{pirelation}). Denoting the images of
$a$ and $b$ respectively by $A$ and~$B$, we thus obtain
\begin{equation}
\hbox{Hom}(\pi,\io) =
\{ \, (A,B)\in \io\times\io \mid BABA^{-1}=E \, \}
\ \ ,
\label{homkb}
\end{equation}
where $E$ stands for the identity element in $\io$.
$\hbox{Hom}(\pi,\io)$ and $\miso$ clearly inherit a topology and
(where smooth) a differentiable structure from those of
$\io\times\io$.

For reasons to be explored in Section~\ref{sec:metrics}, we shall be
mainly interested in those components of $\hbox{Hom}(\pi,\io)$
where $A$ is either in $\iop$ or $\iot$, and $B$ is either in $\ioc$
or $\iotp$ (see Appendix \ref{app:io} for the notation).
As $a$ and $b$ are respectively orientation reversing and
orientation preserving,
these components are precisely the ones that arise from
orientation compatible bundles over $\BbbR\times\KB$. We shall
therefore examine in detail only these four cases and, for the sake
of contrast, the case where both $A$ and $B$ are in $\ioc$. We devote
a separate subsection to each case.

\subsection{$\moo$}
\label{subsec:trivbundle}

We begin by taking $A\in\ioc$ and $B\in\ioc$.
This part of~$\miso$ is
\begin{equation}
\moo =
\hbox{Hom}(\pi,\ioc)/\ioc
\ \ ,
\label{moo}
\end{equation}
where
\begin{equation}
\hbox{Hom}(\pi,\ioc) =
\{ \, (A,B)\in \ioc\times\ioc \mid BABA^{-1}=E \, \}
\ \ .
\label{homc}
\end{equation}

Analyzing $\moo$ is straightforward. In the notation of
Appendix~\ref{app:io}, we write $A=(R_A,w_A)$ and $B=(R_B,w_B)$. One
way to proceed is to use the parametrization $R_B=\exp(v^IK_I)$, in
terms of which the $\oc$ component of the relation $BABA^{-1}=E$
takes the more transparent form $\exp(v^IK_I)
\exp[{(R_Av)}^IK_I]=\openone$. We shall now list the points in $\moo$
by giving for each point a unique representative in
$\hbox{Hom}(\pi,\ioc)$~(\ref{homc}). The parameters take arbitrary
values except when otherwise stated. There are seven different
subsets, denoted by $A_1$ to~$A_7$.

For $A_1$,
\begin{equation}
R_A =
\exp(\mu K_2)
\ \ , \ \
w_A =
\left(
\begin{array}{c}
0 \\
0 \\
b
\end{array}
\right)
\ \ , \ \
B = (\openone, 0)
\ \ ,
\label{a1}
\end{equation}
where $\mu>0$.

For $A_2$,
\begin{equation}
R_A =
\exp \left( \pm K_0 + K_2 \right)
\ \ , \ \
w_A =
\casehalf p
\left(
\begin{array}{c}
\mp 1 \\
0 \\
1
\end{array}
\right)
\ \ , \ \
B = (\openone, 0)
\ \ .
\label{a2}
\end{equation}
Here, and from now on everywhere with $\pm$ signs, upper and lower
signs give distinct sets of points.

For $A_3$,
\begin{equation}
\begin{array}{rcl}
&&R_A=\openone
\ \ , \ \
B = (\openone, 0)
\ \ ,
\\
\noalign{\bigskip}
&&w_A = \left(
\begin{array}{c}
{\tilde b} \\
0 \\
0
\end{array}
\right)
\ \ {\rm or} \ \
w_A = \left(
\begin{array}{c}
\pm 1 \\
0 \\
1
\end{array}
\right)
\ \ {\rm or} \ \
w_A = \left(
\begin{array}{c}
0 \\
0 \\
b
\end{array}
\right)
\ \ ,
\end{array}
\label{a3}
\end{equation}
where $b>0$. The three possibilities for $w_A$ label the $\oc$
conjugacy classes of timelike or zero, null nonzero, and spacelike
vectors in~$\mink$.

For $A_4$,
\begin{equation}
R_A =
\exp({\tilde\mu} K_0)
\ \ , \ \
w_A =
\left(
\begin{array}{c}
- {\tilde b} \\
0 \\
0
\end{array}
\right)
\ \ , \ \
B = (\openone, 0)
\ \ ,
\label{a4}
\end{equation}
where $0<{\tilde\mu}<\pi$ or
$\pi<{\tilde\mu}<2\pi$.

For $A_5$,
\begin{equation}
R_A =
\exp(\pi K_0)
\ \ , \ \
w_A =
\left(
\begin{array}{c}
- {\tilde b} \\
0 \\
0
\end{array}
\right)
\ \ , \ \
R_B = \openone
\ \ , \ \
w_B =
\left(
\begin{array}{c}
0 \\
0 \\
a
\end{array}
\right)
\ \ ,
\label{a5}
\end{equation}
where $a\geq0$.

For $A_6$,
\begin{equation}
R_A =
\exp(\pi K_0)
\ \ , \ \
w_A = 0
\ \ , \ \
R_B =
\exp(\lambda K_2)
\ \ , \ \
w_B =
\left(
\begin{array}{c}
0 \\
0 \\
a
\end{array}
\right)
\ \ ,
\label{a6}
\end{equation}
where $\lambda>0$.

For $A_7$,
\begin{equation}
R_A =
\exp({\tilde\mu} K_0)
\ \ , \ \
w_A =
\left(
\begin{array}{c}
- {\tilde b} \\
0 \\
0
\end{array}
\right)
\ \ , \ \
R_B = \exp(\pi K_0)
\ \ , \ \
w_B = 0
\ \ ,
\label{a7}
\end{equation}
where $0\leq{\tilde\mu}<2\pi$.

We see that $\moo$ consists of two connected components, one given by
$\bigcup_{i=1}^6 A_i$ and the other by~$A_7$. The latter component is
clearly a manifold with topology $S^1\times \BbbR$. The former
component is close to being a two-dimensional
manifold but contains certain singular subsets. The projection of
$\bigcup_{i=1}^6 A_i$ into $\hbox{Hom}(\pi,\oc)/\oc$ is shown in
Figure~\ref{fig:moo}.

To examine connections that give rise to the points in~$\moo$, we
envisage $\KB$ as the closed square ${\overline Q}=\left\{(x,y)
\mid 0\le x \le 1,\; 0\le y \le 1\right\}$ with the boundaries
identified in the manner explained in Appendix~\ref{app:kb}. In the
coordinate patch consisting of the open square $Q=\left\{(x,y) \mid
0 < x < 1,\; 0 < y < 1\right\}$,
define the one-forms
\begin{equation}
A^0 = {\tilde\mu} dy
\ \ , \ \
e^0 = -{\tilde b} dy
\ \ ,
\label{a4c}
\end{equation}
where ${\tilde\mu}$ and ${\tilde b}$ are constants. It is clear that
these one-forms uniquely continue into smooth one-forms on~$\KB$.
The continued one-forms define a flat connection on the
trivial principal $\ioc$ bundle over~$\BbbR\times\KB$,
and the holonomies of
this connection are given by~(\ref{a4}). It is straightforward to
find one-forms on $\KB$ with similar properties also for the
holonomies (\ref{a1})--(\ref{a3}).

To understand the holonomies (\ref{a5}) and~(\ref{a6}), it is easiest
to introduce a local chart $(\BbbR\times Q)\times\ioc$ in which the
transition function upon exiting and re-entering across the vertical
boundaries $x=0$ and $x=1$ is still the identity, but the transition
function upon exiting and re-entering across the horizontal
boundaries $y=0$ and $y=1$ is $(\exp(\pi K_0),0)$. It is clear that
this is a local chart in the trivial principal $\ioc$ bundle
over~$\BbbR\times\KB$. In this chart, consider the one-forms
\begin{equation}
A^2 = \lambda dx
\ \ , \ \
e^2 = b dx
\ \ ,
\label{a6c}
\end{equation}
where $\lambda$ and $b$ are constants. These one-forms define a flat
connection on the trivial principal $\ioc$ bundle
over~$\BbbR\times\KB$, and the
holonomies of this connection are given by~(\ref{a6}). In a global
chart, one-forms gauge equivalent to (\ref{a6c}) are given by
\begin{equation}
\begin{array}{rcl}
&&A^0 = \pi dy
\ \ , \ \
A^1 = \sin(\pi y) \lambda dx
\ \ , \ \
A^2 = \cos(\pi y) \lambda dx
\ \ ,
\\
\noalign{\smallskip}
&&
e^1 = \sin(\pi y) b dx
\ \ , \ \
e^2 = \cos(\pi y) b dx
\ \ .
\end{array}
\label{a6cp}
\end{equation}
Finding one-forms with the analogous properties for the holonomies
(\ref{a5}) is straightforward.

To understand the remaining component~$A_7$, consider again the
one-forms on $\KB$ defined by~(\ref{a4c}). Consider the $\ioc$ bundle
over $\BbbR\times\KB$
such that there exists a local chart $(\BbbR\times Q)\times\ioc$
in which the transition function
across the horizontal boundaries
$y=0$ and $y=1$ is the identity,
but the transition function
across the vertical boundaries $x=0$ and $x=1$
is $(\exp(\pi K_0),0)$. It is
straightforward to verify that this defines a nontrivial bundle whose
Euler characteristic\cite{amano} is equal to~1. The local expressions
(\ref{a4c}) define by continuity a flat connection on this bundle,
and the holonomies of this flat connection are given by~(\ref{a7}).

\subsection{$\mto$}
\label{subsec:Tbundle}

We next take $A\in\iot$ and $B\in\ioc$.
This part of~$\miso$ is
\begin{equation}
\mto =
\{ \, (A,B)\in \iot\times\ioc \mid BABA^{-1}=E \, \}/\ioc
\ \ .
\label{mto}
\end{equation}

We can proceed as with~$\moo$. Writing $R_A=-R$, where $R\in\oc$,
one immediately sees that the projection of $\mto$ into
$\hbox{Hom}(\pi,\o)/\oc$ is
homeomorphic to the corresponding
projection of~$\moo$. We shall therefore divide $\mto$ into
seven different subsets, denoted by $B_1$ to~$B_7$,
whose projections into $\hbox{Hom}(\pi,\o)/\oc$ are
homeomorphic to the corresponding projections of the sets $A_1$
to~$A_7$. As before, the parameters take arbitrary values except when
otherwise stated, and the parametrization is unique.

For $B_1$,
\begin{equation}
R_A =
-\exp(\mu K_2)
\ \ , \ \
w_A = 0
\ \ , \ \
R_B = \openone
\ \ , \ \
w_B =
\left(
\begin{array}{c}
0 \\
0 \\
a
\end{array}
\right)
\label{b1}
\end{equation}
where $\mu>0$.

For $B_2$,
\begin{equation}
R_A =
- \exp \left( \pm K_0+K_2 \right)
\ \ , \ \
w_A = 0
\ \ , \ \
R_B = \openone
\ \ , \ \
w_B =
\casehalf p
\left(
\begin{array}{c}
\pm1 \\
0 \\
1
\end{array}
\right)
\ \ .
\label{b2}
\end{equation}

For $B_3$,
\begin{equation}
\begin{array}{rcl}
&&A = (-\openone,0)
\ \ , \ \
R_B=\openone
\ \ ,
\\
\noalign{\bigskip}
&&w_B = \left(
\begin{array}{c}
{\tilde a} \\
0 \\
0
\end{array}
\right)
\ \ {\rm or} \ \
w_B = \left(
\begin{array}{c}
\pm1 \\
0 \\
1
\end{array}
\right)
\ \ {\rm or} \ \
w_B = \left(
\begin{array}{c}
0 \\
0 \\
a
\end{array}
\right)
\ \ ,
\end{array}
\label{b3}
\end{equation}
where $a>0$.

For $B_4$,
\begin{equation}
R_A =
-\exp({\tilde\mu} K_0)
\ \ , \ \
w_A = 0
\ \ , \ \
R_B = \openone
\ \ , \ \
w_B =
\left(
\begin{array}{c}
- {\tilde a} \\
0 \\
0
\end{array}
\right)
\ \ ,
\label{b4}
\end{equation}
where $0<{\tilde\mu}<\pi$ or $\pi<{\tilde\mu}<2\pi$.

For $B_5$,
\begin{equation}
R_A =
-\exp(\pi K_0)
\ \ , \ \
w_A =
\left(
\begin{array}{c}
0 \\
0 \\
b
\end{array}
\right)
\ \ , \ \
R_B = \openone
\ \ , \ \
w_B =
\left(
\begin{array}{c}
- {\tilde a} \\
0 \\
0
\end{array}
\right)
\ \ ,
\label{b5}
\end{equation}
where $b\geq0$.

For $B_6$,
\begin{equation}
R_A =
-\exp(\pi K_0)
\ \ , \ \
w_A =
\left(
\begin{array}{c}
0 \\
0 \\
b
\end{array}
\right)
\ \ , \ \
R_B =
\exp(\lambda K_2)
\ \ , \ \
w_B = 0
\ \ ,
\label{b6}
\end{equation}
where $\lambda>0$.

For $B_7$,
\begin{equation}
R_A =
-\exp({\tilde\mu} K_0)
\ \ , \ \
w_A = 0
\ \ , \ \
R_B = \exp(\pi K_0)
\ \ , \ \
w_B =
\left(
\begin{array}{c}
- {\tilde a} \\
0 \\
0
\end{array}
\right)
\ \ ,
\label{b7}
\end{equation}
where $0\leq{\tilde\mu}<2\pi$.

We see that $\mto$ consists of two connected components, one given by
$\bigcup_{i=1}^6 B_i$ and the other by~$B_7$. They are closely
analogous to the corresponding components of~$\moo$.

Connections that yield the points in $\mto$ can again be
investigated by envisaging $\KB$ as the closed square~${\overline Q}$
and constructing the bundle in terms of a local chart
associated with the open square~$Q$. The component $\bigcup_{i=1}^6
B_i$ comes from a bundle that admits charts
in which the transition function
across the vertical boundaries $x=0$ and $x=1$
is the identity, but the transition function
across the horizontal boundaries
$y=0$ and $y=1$ is $(-\openone,0)$.
The component $B_7$ comes from a bundle
in which the transition function across the horizontal
boundaries is as above but the transition function across
the vertical boundaries is $(\exp(\pi K_0),0)$.
Expressions for the components of the connection
one-form in these local charts are easily written down;
for example, the one-form
\begin{equation}
A^0 = {\tilde\mu} dy
\ \ , \ \
e^0 = - {\tilde a} dx
\label{b4c}
\end{equation}
gives the holonomies $B_4$ (\ref{b4}) and $B_7$ (\ref{b7}) in the
appropriate bundles.

\subsection{$\mpo$}
\label{subsec:Pbundle}

We next take $A\in\iop$ and $B\in\ioc$.
This part of~$\miso$ is
\begin{equation}
\mpo =
\{ \, (A,B)\in \iop\times\ioc \mid BABA^{-1}=E \, \}/\ioc
\ \ .
\label{mpo}
\end{equation}

We can proceed as above. Analyzing the conjugacy classes is now
more involved than in the previous cases, and we outline in
Appendix \ref{app:mmpo} this analysis for the $\o$ part of~$\mpo$.
$\mpo$ consists of five different subsets, denoted by $C_1$ to~$C_5$.
As before, the parameters take arbitrary values except when otherwise
stated, and the parametrization is unique. We write $P={\rm
diag}(1,-1,1)$.

For $C_1$,
\begin{equation}
R_A = P
\ \ , \ \
w_A =
\left(
\begin{array}{c}
0 \\
0 \\
b
\end{array}
\right)
\ \ , \ \
R_B =
\exp(\lambda K_2)
\ \ , \ \
w_B = 0
\ \ ,
\label{c1}
\end{equation}
where $\lambda>0$.

For $C_2$,
\begin{equation}
R_A = P
\ \ , \ \
w_A =
\casehalf p
\left(
\begin{array}{c}
\pm 1 \\
0 \\
1
\end{array}
\right)
\ \ , \ \
R_B =
\exp ( \pm K_0+K_2 )
\ \ , \ \
w_B = 0
\ \ .
\label{c2}
\end{equation}

For $C_3$,
\begin{equation}
\begin{array}{rcl}
&&R_A = P
\ \ , \ \
R_B = \openone
\ \ , \ \
w_B =
\left(
\begin{array}{c}
0 \\
a \\
0
\end{array}
\right)
\ \ , \ \
\\
\noalign{\bigskip}
&&w_A =
\left(
\begin{array}{c}
{\tilde b} \\
0 \\
0
\end{array}
\right)
\ \ {\rm or} \ \
w_A = \left(
\begin{array}{c}
\pm1 \\
0 \\
1
\end{array}
\right)
\ \ {\rm or} \ \
w_A = \left(
\begin{array}{c}
0 \\
0 \\
b
\end{array}
\right)
\ \ ,
\end{array}
\label{c3}
\end{equation}
where $b>0$, and in addition $a\ge0$ when $w_A$ is given by the first
of the three alternatives.

For $C_4$,
\begin{equation}
R_A = P
\ \ , \ \
w_A =
\left(
\begin{array}{c}
- {\tilde b} \\
0 \\
0
\end{array}
\right)
\ \ , \ \
R_B =
\exp({\tilde\lambda} K_0)
\ \ , \ \
w_B = 0
\ \ ,
\label{c4}
\end{equation}
where $0<{\tilde\lambda}<2\pi$.

For $C_5$,
\begin{equation}
R_A = P \exp(\mu K_1)
\ \ , \ \
w_A = 0
\ \ , \ \
R_B = \openone
\ \ , \ \
w_B =
\left(
\begin{array}{c}
0 \\
a \\
0
\end{array}
\right)
\ \ ,
\label{c5}
\end{equation}
where $\mu>0$.

We see that $\mpo$ is connected, and close to being a two dimensional
manifold. The projection of $\mpo$ into $\hbox{Hom}(\pi,\o)/\oc$
is shown in Figure~\ref{fig:mpo}. In terms of
a local chart associated with the square ${\overline Q}$
as before, the bundle admits
charts in which the transition function
across the vertical
boundaries $x=0$ and $x=1$ is the identity, but
the transition function
across the horizontal boundaries $y=0$
and $y=1$ is $(P,0)$. Expressions for
the components of the connection one-form in such a
chart are again easily found.
For future reference, we list them all here:
\begin{mathletters}
\label{c-all-c}
\begin{eqnarray}
C_1:&&\ \
A^2 = \lambda dx
\ \ , \ \
e^2 = b dy
\ \ ;
\label{c1c}
\\
C_2:&&\ \
\pm A^0 = A^2 = dx
\ \ , \ \
\pm e^0 = e^2 = \casehalf p dy
\ \ ;
\label{c2c}
\\
C_3:&&\ \
A^I = 0
\ \ , \ \
e^1 = a dx
\ \ , \ \
\nonumber
\\
&&\ \ \ \
e^0 = {\tilde b} dy
\ \ \hbox{or} \ \
\pm e^0 = e^2 = dy
\ \ \hbox{or} \ \
e^2 = b dy
\ \ ;
\label{c3c}
\\
C_4:&&\ \
A^0 = {\tilde\lambda} dx
\ \ , \ \
e^0 = -{\tilde b} dy
\ \ ;
\label{c4c}
\\
C_5:&&\ \
A^1 = \mu dy
\ \ , \ \
e^1 = a dx
\ \ .
\label{c5c}
\end{eqnarray}
\end{mathletters}%

\subsection{$\mptwo$}
\label{subsec:PTbundle1}

We next take $A\in\iop$ and $B\in\iotp$.
This part of $\miso$ is
\begin{equation}
\mptwo =
\{ \, (A,B)\in \iop\times\iotp \mid BABA^{-1}=E \, \}/\ioc
\ \ .
\label{mptwo}
\end{equation}

We can again proceed as above. Analyzing the conjugacy classes is now
even more involved; one way to proceed is to use the results given at
the end of Appendix \ref{app:mmpo} about the $\oc$ conjugacy classes
in $\op$. One finds that $\mptwo$ consists of two subsets, which we
denote by $D_1$ and~$D_2$. Representatives for points in these
subsets are given respectively by
\begin{equation}
R_A = P
\ \ , \ \
w_A =
\left(
\begin{array}{c}
0 \\
0 \\
b
\end{array}
\right)
\ \ , \ \
R_B =
-P \exp(\pi K_0) \exp (\lambda K_2)
\ \ , \ \
w_B = 0
\ \ ,
\label{d1}
\end{equation}
and
\begin{equation}
R_A = P \exp(\mu K_1)
\ \ , \ \
w_A = 0
\ \ , \ \
R_B = -P
\ \ , \ \
w_B =
\left(
\begin{array}{c}
0 \\
a \\
0
\end{array}
\right)
\ \ ,
\label{d2}
\end{equation}
and the parametrizations become unique after the identifications
$(\lambda,b)\sim(-\lambda,-b)$ and $(\mu,a)\sim(-\mu,-a)$.

The connected components of $\mptwo$ are just $D_1$ and~$D_2$. Each
becomes a two-dimensional manifold if the points $\lambda=b=0$ and
$\mu=a=0$ are respectively excised. In terms of a local chart
associated with the
square~${\overline Q}$ as before, the bundle corresponding to $D_1$
admits charts in which the transition function across the
vertical boundaries $x=0$ and $x=1$ is $(-P\exp(\pi K_0),0)$,
whereas the transition function
across the horizontal boundaries $y=0$ and $y=1$ is $(P,0)$.
In the bundle corresponding to $D_2$ the transition function
across the horizontal boundaries is as above but
the transition function across the vertical
boundaries is $(-P,0)$. Expressions for
the components of the connection one-form in such
charts are easily found: for $D_1$ and $D_2$ one has
respectively
\begin{equation}
A^2 =  \lambda dx
\ \ , \ \
e^2 = b dy
\ \ ,
\label{d1c}
\end{equation}
and
\begin{equation}
A^1 =  \mu dy
\ \ , \ \
e^1 = a dx
\ \ .
\label{d2c}
\end{equation}

\subsection{$\mttwo$}
\label{subsec:PTbundle2}

We finally take $A\in\iot$ and $B\in\iotp$.
This part of $\miso$ is
\begin{equation}
\mttwo =
\{ \, (A,B)\in \iot\times\iotp \mid BABA^{-1}=E \, \}/\ioc
\ \ .
\label{mttwo}
\end{equation}
$\mttwo$ is isomorphic to $\mptwo$ via the
map $(A,B) \mapsto (AB,B)$.
In the notation of Appendix~\ref{app:kb}, this
isomorphism is generated by the element
$(1, -1)$ of ${\rm Out}(\pi) \simeq \BbbZ_2 \times \BbbZ_2$.

\section{Spacetime metrics on ${\BbbR} \times\hbox{(Klein bottle)}$}
\label{sec:metrics}

In this section we shall examine nondegenerate metrics that are
recovered from the connection solutions.
We shall show that four of the components of $\miso$ contain points
that yield such metrics.

\subsection{$\mpo$}
\label{subsec:mPbundle}

We begin with the component~$\mpo$.

Consider first the set $C_1$~(\ref{c1}). In a local chart
over the set $\BbbR\times Q\subset \BbbR\times\KB$ defined as in
subsection~\ref{subsec:Pbundle}, a connection one-form giving the
holonomies (\ref{c1}) is given by~(\ref{c1c}).
Changing the local chart by the transition function
$(\openone, {(t,0,0)}^T)$, where $t$ is the coordinate on~$\BbbR$,
$A^I$ remains unchanged but the new triad is
\begin{equation}
e^0 = dt
\ \ , \ \
e^1 = t\lambda dx
\ \ , \ \
e^2 = b dy
\ \ .
\label{c1ftriad1}
\end{equation}
By assumption $\lambda>0$. If now $b\neq0$, squaring
(\ref{c1ftriad1}) gives on $(0,\infty)\times Q$ the nondegenerate
metric
\begin{equation}
ds^2 = -dt^2 + t^2 \lambda^2 dx^2 + b^2 dy^2
\ \ ,
\label{c1fmetric1}
\end{equation}
and adding charts with $\o$ transition functions to cover
$(0,\infty)\times\KB$ as explained in subsection
\ref{subsec:kinematics} clearly gives on $(0,\infty)\times\KB$ the
metric obtained by continuation of~(\ref{c1fmetric1}). From the
identifications (\ref{abmap}) in Appendix \ref{app:kb} one sees that
this spacetime is constructed by taking the quotient of the
region $T>|X|$ in $\mink$ with respect to the two
holonomies~(\ref{c1}).
This spacetime is a modest generalization of one causal
region of Misner space\cite{haw-ell}.

Consider then the set $C_5$~(\ref{c5}). Proceeding as above, we
start from the connection one-form~(\ref{c5c}), and change
the local chart by $(\openone, {(t,0,0)}^T)$ to
obtain the new triad
\begin{equation}
e^0 = dt
\ \ , \ \
e^1 = a dx
\ \ , \ \
e^2 = -t\mu dy
\ \ .
\label{c5ftriad1}
\end{equation}
By assumption $\mu>0$. If now $a\neq0$, one obtains as
above on $(0,\infty)\times\KB$ the nondegenerate metric
\begin{equation}
ds^2 = -dt^2 + t^2 \mu^2 dy^2 + a^2 dx^2
\ \ .
\label{c5fmetric1}
\end{equation}
The spacetime is constructed by taking the quotient of the region
$T>|Y|$ in $\mink$ with respect to the two holonomies~(\ref{c5}).
This spacetime is therefore another generalization of one causal
region of Misner space\cite{haw-ell}.
Note that it is not isometric to~(\ref{c1fmetric1}).

Consider next the set~$C_3$, with $w_A$ given by the last alternative
in~(\ref{c3}). By assumption $b>0$. If $a\neq0$, one proceeds as
above from (\ref{c3c}) to obtain on
$(-\infty,\infty)\times\KB$ the nondegenerate metric
\begin{equation}
ds^2 = -dt^2 + a^2 dx^2 + b^2 dy^2
\ \ .
\label{c3fmetric1}
\end{equation}
The spacetime is constructed by taking the quotient of $\mink$ with
respect to the two holonomies~(\ref{c3}).

All the three classes of spacetimes (\ref{c1fmetric1}),
(\ref{c5fmetric1}), and (\ref{c3fmetric1}) are globally
hyperbolic, with $\KB$ spacelike surfaces, and (\ref{c3fmetric1}) is
in addition geodesically complete. Further, these spacetimes are
domains of dependence in the sense of Ref.\cite{mess}. It appears
likely that the methods of Refs.\cite{moncrief1,mess,hosoya-nakao1}
could be adapted to show that these three classes of
spacetimes exhaust, up to isometries,
all flat 2+1 spacetimes that are domains of dependence
of a spacelike Klein bottle\cite{mess-priv}.

The spacetimes (\ref{c1fmetric1}), (\ref{c5fmetric1}), and
(\ref{c3fmetric1}) were obtained by taking the quotient of $\mink$ or
some subset of it with respect the two $\io$ holonomies.
This is similar to what happens in the $\ioc$ connection formulation
of 2+1 gravity on $\BbbR\times
T^2$\cite{moncrief1,carlip1,mess,louma}. We shall now show that the
similarity extends further, in that all points of~$\mpo$, except a
set of measure zero, yield nondegenerate spacetime metrics.

Let us first reconsider the set $C_5$ with $a\neq0$. A gauge
transformation of (\ref{c5c}) with the transition function
$(\openone, {(\casehalf (t+1),0,\casehalf (t-1))}^T)$ leads to the
triad
\begin{equation}
\begin{array}{rcl}
e^0 &=& \casehalf dt - \casehalf (t-1) \mu dy
\\
e^1 &=& a dx
\\
e^2 &=& \casehalf dt - \casehalf (t+1) \mu dy
\ \ ,
\end{array}
\label{c1ftriad2}
\end{equation}
which is nondegenerate for $t>0$. It is straightforward to verify,
using the coordinate transformations given in
Ref.\cite{haw-ell} in the context of Misner space, that
the metric arising from (\ref{c1ftriad2}) describes the spacetime
that is obtained as the quotient of the region $T>-Y$
in $\mink$ with respect to
the two holonomies~(\ref{c5}). The $t={\rm constant}$ Klein bottles
go from timelike, for $t<0$, via null, at $t=0$, to spacelike, for
$t>0$.

For $C_3$ the situation is straightforward. Whenever $w_A$ and $w_B$
in (\ref{c3}) are linearly independent, one easily finds a
nondegenerate co-triad such that the resulting spacetime is simply
the quotient of $\mink$ with respect to the two holonomies. The Klein
bottles are spacelike only in the case~(\ref{c3fmetric1}).

Consider then the set~$C_2$. Starting from the connection one-form
(\ref{c2c}) and performing a gauge transformation by the transition
function $(\openone, {(\pm\casehalf t, 0, -\casehalf t)}^T)$ leads to
the triad
\begin{equation}
\begin{array}{rcl}
e^0 &=& \pm\casehalf p dy \pm \casehalf dt
\\
e^1 &=& \pm t dx
\\
e^2 &=& \casehalf p dy - \casehalf dt
\ \ .
\end{array}
\label{c2ftriad}
\end{equation}
If $p\ne0$, this triad is is nondegenerate for $t>0$. The spacetime
is obtained as the quotient of the region $T>\pm Y$ in $\mink$ with
respect to the holonomies~(\ref{c2}).

Finally, consider the set~$C_4$. Performing on (\ref{c4c}) a gauge
transformation by the transition function $(\openone,
{(0,0,t)}^T)$ leads to the triad
\begin{equation}
e^0 = -{\tilde b} dy
\ \ , \ \
e^1 = - t {\tilde\lambda} dx
\ \ , \ \
e^2 = dt
\ \ .
\label{c4ftriad}
\end{equation}
By assumption ${\tilde\lambda}>0$. If ${\tilde b}\ne0$, the triad
(\ref{c4ftriad}) is nondegenerate for $t>0$, and the metric is
\begin{equation}
ds^2 = - {\tilde b}^2 dy^2
+ dt^2
+ t^2 {\tilde\lambda}^2 dx^2
\ \ .
\label{c4fmetric}
\end{equation}
This spacetime can be obtained in the following way. One first
removes from $\mink$ the axis $X=Y=0$ and passes to the angular
coordinates $(T,\rho,\varphi)$ by
\begin{equation}
X = \rho \sin\varphi
\ \ , \ \
Y = - \rho \cos\varphi
\ \ ,
\end{equation}
where $\rho>0$ and $\varphi$ is identified with period~$2\pi$. One
then passes to the universal covering space, which means dropping the
periodicity of~$\varphi$.
Finally, one takes the quotient of this covering space with respect
to the two isometries
\begin{mathletters}
\label{c4isometries}
\begin{eqnarray}
&&(T,\rho,\varphi) \mapsto (T- {\tilde b},\rho,-\varphi)
\ \ ,
\\
&&(T,\rho,\varphi) \mapsto (T,\rho,\varphi + {\tilde\lambda})
\ \ .
\end{eqnarray}
\end{mathletters}%
The metric (\ref{c4fmetric}) is obtained through the coordinate
transformation
\begin{equation}
\begin{array}{rcl}
T &=& {\tilde\lambda}y
\ \ ,
\\
\rho &=& t
\ \ ,
\\
\varphi &=& {\tilde \lambda} x
\ \ .
\end{array}
\end{equation}
If $\varphi$ were periodic with period~$2\pi$, the isometries
(\ref{c4isometries}) would be precisely the holonomies~(\ref{c4}).
However, when $\varphi$ is not periodic, the holonomies~(\ref{c4}) do
not have a natural (associative) action for generic values
of~${\tilde\lambda}$. This means that the holonomies (\ref{c4})
represent the isometries (\ref{c4isometries})
of the covering space only in a local
sense.\footnote{This point is missed in the corresponding discussion
for $\BbbR\times T^2$ in Ref.\cite{louma}. The statement therein that
the $\ioc$ holonomies can be used for quotienting the covering space
is incorrect.}

Triads gauge-equivalent to (\ref{c4ftriad}) are obtained by adding
multiples of $2\pi$ to~${\tilde\lambda}$. For generic values of
${\tilde\lambda}$, the resulting metrics are not diffeomorphic
to~(\ref{c4fmetric}). A~similar phenomenon was noted for $\BbbR\times
T^2$ in Ref.\cite{louma}.

Finally, note that in all the new local charts that we
obtained through the gauge transformations in this subsection, the
transition functions across the boundaries of $Q$ are the same as in
the original local chart. Had we for example attempted to
adapt the construction of the triad (\ref{c1ftriad2}) to the
set~$C_1$, with the transition function $(\openone, {(\casehalf
(t+1),\casehalf (t-1))}^T,0)$, this would no longer have been true:
the new transition function across the horizontal boundaries of $Q$
would not have been in $\o$, and the metrics across the boundary
would not have agreed. This is related to the fact that the
holonomies (\ref{c5}) preserve the domain $T>-Y$ in~$\mink$, but the
holonomies (\ref{c1}) do not preserve the domain $T>-X$.

\subsection{$\mto$}
\label{subsec:mTbundle}

We now turn to~$\mto$.

Let us begin with the sets $B_4$, $B_5$, and~$B_6$.
It is most convenient not to use the local chart mentioned in
subsection~\ref{subsec:Tbundle}, but instead a chart in
which the transition function across the vertical boundaries $x=0$
and $x=1$ is the identity, and the transition function
across the horizontal boundaries
$y=0$ and $y=1$ is $(-\exp(\pi K_0),0)$. Connection
one-forms for $B_4$, $B_5$, and $B_6$
in this chart are given respectively by
\begin{equation}
A^0 = ({\tilde\mu} + \pi) dy
\ \ , \ \
e^0 = - {\tilde a} dx
\ \ , \ \
e^1 = - ({\tilde\mu} + \pi)t dy
\ \ , \ \
e^2 = dt
\ \ ,
\label{b4c2}
\end{equation}
\begin{equation}
A^I = 0
\ \ , \ \
e^0 = - {\tilde a} dx
\ \ , \ \
e^1 = dt
\ \ , \ \
e^2 = b dy
\ \ ,
\label{b5c2}
\end{equation}
and
\begin{equation}
A^2 = \lambda dx
\ \ , \ \
e^0 = \lambda t dx
\ \ , \ \
e^1 = dt
\ \ , \ \
e^2 = b dy
\ \ .
\label{b6c2}
\end{equation}
The metric coming from (\ref{b5c2}) describes,
for ${\tilde a}\ne0\ne b$,
the spacetime obtained as the quotient of $\mink$ with respect to
the holonomies~(\ref{b5}). Similarly, the metric coming from
(\ref{b6c2}) describes, for $\lambda\ne0\ne b$, the spacetime
obtained by taking the quotient of the Rindler wedge $X>|T|$ with
respect to the holonomies~(\ref{b6}).
The metric coming from (\ref{b4c2})
describes, for ${\tilde\mu} + \pi\ne0\ne {\tilde a}$, the spacetime
obtained by first cutting out the $T$-axis from~$\mink$, then going
to the universal covering space, and finally taking a quotient with
respect to two isometries that are locally represented
by the holonomies~(\ref{b4}).
All these spacetimes are space
orientable but time-nonorientable. Again, connection one-forms
gauge-equivalent to (\ref{b4c2}) are obtained by adding multiples of
$2\pi$ to~${\tilde\lambda}$.

Let us next consider~$B_1$. In the local chart mentioned in
subsection~\ref{subsec:Tbundle}, a connection one-form yielding the
holonomies (\ref{b1}) is
\begin{equation}
A^2 = \mu dy
\ \ , \ \
e^2 = a dx
\ \ .
\label{b1c}
\end{equation}
We now change the chart by the transition function
$(\openone, {(t \cos(\pi y),t \sin(\pi y),0)}^T)$. The new triad is
\begin{equation}
\begin{array}{rcl}
e^0 &=& \cos(\pi y) dt + t (\mu-\pi) \sin(\pi y) dy
\ \ ,
\\
e^1 &=& \sin(\pi y) dt + t (\mu+\pi) \cos(\pi y) dy
\ \ ,
\\
e^2 &=& a dx
\ \ ,
\end{array}
\label{b1ftriad1}
\end{equation}
and in the new chart the transition functions across the
boundaries are the same as in the old chart. Squaring the
triad (\ref{b1ftriad1}) gives the metric
\begin{eqnarray}
ds^2 &=& - \cos(2\pi y) dt^2
+ t^2 \left[ (\mu^2 + \pi^2) \cos(2\pi y)
+ 2\pi \mu \right] dy^2
\nonumber
\\
&&+ 2\pi \sin(2\pi y) t dt dy
+ a^2 dx^2
\ \ .
\label{b1fmetric1}
\end{eqnarray}
Recall that $\mu>0$. From now on we assume
$\mu<\pi$ and $a\ne0$. The metric (\ref{b1fmetric1}) is then
nondegenerate for $t>0$, and it clearly defines a
nondegenerate metric on $\BbbR \times \KB$.

To understand~(\ref{b1fmetric1}), consider for the moment the metric
\begin{eqnarray}
ds^2 &=& - \cos(2\pi \theta) dt^2
+ t^2 \left[ (\mu^2 + \pi^2) \cos(2\pi \theta)
+ 2\pi \mu \right] d\theta^2
\nonumber
\\
&&+ 2\pi \sin(2\pi \theta) t dt d\theta
+ a^2 d\varphi^2
\ \ ,
\label{b1twometric}
\end{eqnarray}
where $t>0$, and $\theta$ and $\varphi$ both take all real values.
Define a transformation to the new coordinates $(U,V,Y)$ by
\begin{equation}
\begin{array}{rcl}
U &=& 2^{1/2} t \exp(\mu \theta) \sin(\pi \theta + \pi/4)
\ \ ,
\\
V &=& 2^{1/2} t \exp(-\mu \theta) \cos(\pi \theta + \pi/4)
\ \ ,
\\
Y &=& a \varphi
\ \ .
\end{array}
\label{bnulltrans}
\end{equation}
The transformation is clearly not globally one-to-one, but in any
sufficiently small interval in $\theta$ it is one-to-one to its
image. In any such interval, the metric (\ref{b1twometric}) takes the
form
\begin{equation}
ds^2 = -dUdV + dY^2
\ \ ,
\label{minknullmetric}
\end{equation}
which is the 2+1 Minkowski metric in double null coordinates.
Patching together the intervals in~$\theta$, we thus see that the
metric~(\ref{b1twometric})
describes the space $\tilde B$ that is obtained by removing from
$\mink$ a spacelike geodesic and then passing to the universal
covering space. We can think of the system $(U,V,Y)$ as a set of
coordinates on a single sheet of~$\tilde B$, with the removed
geodesic being at $U=V=0$. With an appropriately placed cut in the
system $(U,V,Y)$, the isometry
\begin{mathletters}
\label{Btilde-isom}
\begin{equation}
(t,\varphi,\theta) \mapsto (t,-\varphi,\theta+1)
\end{equation}
has the effect $(U,V,Y) \mapsto (- e^\mu U, - e^{-\mu} V,-Y)$:
this is a boost in the constant $Y$ planes with rapidity~$\mu$,
followed by the inversion $(U,V,Y) \mapsto (-U,-V,-Y)$. The isometry
\begin{equation}
(t,\varphi,\theta) \mapsto (t,\varphi+1,\theta)
\end{equation}
\end{mathletters}%
is a translation in $Y$ with
magnitude~$a$. Comparing (\ref{b1fmetric1}) to~(\ref{b1twometric}),
and recalling the identifications of the coordinates $(x,y)$, it
becomes clear that (\ref{b1fmetric1}) is obtained by taking the
quotient of $\tilde B$ with respect to the two
isometries~(\ref{Btilde-isom}),
which are locally represented by the
holonomies~(\ref{b1}).

Finally, let us consider~$B_3$. When $w_A$ is given by the last of
the three alternatives in~(\ref{b3}), a nondegenerate metric is
obtained as the $\mu\to0$ limit of~(\ref{b1fmetric1}). To understand
this spacetime as a quotient space, it is now not necessary to pass
to the covering space after removing the spacelike geodesic
from~$\mink$: the spacetime is simply obtained by removing from
$\mink$ the $Y$ axis and taking the quotient with respect to the two
holonomies~(\ref{b3}). It is clear how to adapt the construction to
the case where $w_A$ is given by the second or the first (with
${\tilde a}\ne0$) of the three alternatives in~(\ref{b3}).

\subsection{$\mptwo$ and $\mttwo$}
\label{subsec:mPTbundle}

We finally turn to $\mptwo$ and~$\mttwo$.
We shall demonstrate that in the component
$D_1\subset\mptwo$ (\ref{d1})
there are points that yield
nondegenerate metrics. The same then also holds for the
component of $\mttwo$ that is isomorphic to $D_1$ via an outer
automorphism of~$\pi$.

We start from the connection (\ref{d1c}) in the local chart
mentioned in subsection~\ref{subsec:PTbundle1}, and we change the
chart by the transition function
$(\openone, {(t \cos(\pi x),t \sin(\pi x),0)}^T)$.
The new triad is
\begin{equation}
\begin{array}{rcl}
e^0 &=& \cos(\pi x) dt + t (\lambda-\pi) \sin(\pi x) dx
\ \ ,
\\
e^1 &=& \sin(\pi x) dt + t (\lambda+\pi) \cos(\pi x) dx
\ \ ,
\\
e^2 &=& b dy
\ \ ,
\end{array}
\label{d1ftriad1}
\end{equation}
and in the new chart the transition functions across the
boundaries are the same as in the old chart. Squaring the
triad (\ref{d1ftriad1}) gives the metric
\begin{eqnarray}
ds^2 &=& - \cos(2\pi x) dt^2
+ t^2 \left[ (\lambda^2 + \pi^2) \cos(2\pi x)
+ 2\pi \lambda \right] dx^2
\nonumber
\\
&&+ 2\pi \sin(2\pi x) t dt dx
+ b^2 dy^2
\ \ .
\label{d1fmetric1}
\end{eqnarray}
For $|\lambda|<\pi$ and $b\ne0$, the metric (\ref{d1fmetric1})
is nondegenerate for $t>0$, and it clearly defines a nondegenerate
metric on $\BbbR \times \KB$. Comparing (\ref{d1fmetric1})
with~(\ref{b1twometric}), it is seen as in subsection
\ref{subsec:mTbundle} that the spacetime described by
(\ref{d1fmetric1}) is obtained by taking the quotient of the space
$\tilde B$ with respect to two isometries. In the local null
coordinates $(U,V,Y)$ on~$\tilde B$, one of the isometries is
$(U,V,Y) \mapsto (- e^\lambda U, - e^{-\lambda} V,Y)$: this is a
boost in the constant $Y$ planes with rapidity~$\lambda$, followed by
the space and time inversion in the constant $Y$ planes. The other
isometry is $(U,V,Y) \mapsto (V,U,Y+b)$, which is a space inversion
in the constant $Y$ planes followed by a translation in~$Y$. In local
coordinate patches these isometries are represented by the
holonomies~(\ref{d1}), but globally this is true only when
$\lambda=0$.

\section{Symplectic structure and quantization}
\label{sec:quantization}

In the previous two sections we have investigated several components
of the solution space~$\miso$. In this section we shall briefly
discuss the possibilities for quantizing the theory.

As was mentioned in Section~\ref{sec:kbconnsol}, the spaces $\mto$,
$\mpo$, $\mttwo$, and $\mptwo$ come from orientation compatible
bundles over $\BbbR \times \KB$. This raises the possibility of
endowing (smooth subsets of) these spaces with a symplectic
structure, by first performing a Hamiltonian decomposition of the
action (\ref{action}) as in the orientable
case\cite{witten1,AAbook2,romano} and then pulling the resulting
symplectic structure on the fields back to a symplectic structure
$\Omega$ on the solution space. Using our explicit
parametrizations, it is straightforward to verify that
$\Omega$ is nondegenerate, with the exception of certain subsets
whose projections into $\hbox{Hom}(\pi,\o)/\oc$ have measure zero.
Avenues towards quantization can thus be explored for example via the
geometric quantization techniques of
Refs.\cite{woodhouse,AAbook-geom}.

For brevity, we shall concentrate on~$\mpo$. Analogous considerations
hold for~$\mto$, $\mttwo$, and~$\mptwo$.

The symplectic structure $\Omega$ on (smooth subsets of) $\mpo$ is
readily read off by substituting the connection one-forms
(\ref{c-all-c}) into the Hamiltonian decomposition of the action. The
result is:
\begin{mathletters}
\label{omega}
\begin{eqnarray}
C_1:&&\ \
\Omega =  db \wedge d\lambda
\ \ ;
\label{omegac1}
\\
C_2:&&\ \
\Omega =  0
\ \ ;
\label{omegac2}
\\
C_3:&&\ \
\Omega = 0
\ \ ;
\label{omega3c}
\\
C_4:&&\ \
\Omega =  d{\tilde b} \wedge d{\tilde \lambda}
\ \ ;
\label{omega4c}
\\
C_5:&&\ \
\Omega = - da \wedge d\mu
\ \ .
\label{omega5c}
\end{eqnarray}
\end{mathletters}%
The Poisson brackets are thus
$\left\{ b , \lambda \right\}
= \left\{ {\tilde b} , {\tilde \lambda} \right\}
= - \left\{ a , \mu \right\} = 1$.

Experience with the orientable case\cite{witten1,louma} suggests that
one could use $\Omega$ to interpret most of $\mpo$ as a cotangent
bundle over the subspace where the holonomies are in~$\o$. It is
immediately seen from (\ref{omega}) that this is possible for the
disjoint open sets~$C_1$, $C_4$, and~$C_5$. One can therefore
quantize $C_1$, $C_4$, and $C_5$ in the geometric quantization
framework of Refs.\cite{woodhouse,AAbook-geom}, directly following
the treatment of the $\ioc$ torus theory in Ref.\cite{louma}.
Borrowing terminology that has been invoked for the
torus\cite{AAbook2}, one might refer to the quantum theories arising
from $C_1$ and $C_5$ as two distinct ``spacelike sector" theories,
and to the quantum theory arising from $C_4$ as a ``timelike sector"
theory.

As $\mpo$ is connected, it is natural to ask whether there exist
larger quantum theories that would in some sense connect the
individual quantum theories built from~$C_1$, $C_4$, and~$C_5$. As
$C_5$ is classically connected to $C_1$ and $C_4$ in $\mpo$ only in a
non-smooth manner through~$C_3$, it is not obvious whether there is a
natural larger quantum theory that would connect the quantization of
$C_5$ to the quantizations of $C_1$ or~$C_4$. However, we shall now
show that there is a larger quantum theory that connects the
quantizations of $C_1$ and~$C_4$.

Let us start from $C_1$ in the parametrization~(\ref{c1}), with
$\lambda>0$. We introduce a new parametrization by
\begin{equation}
\begin{array}{rcl}
\lambda &=&
{\displaystyle{
{(2\sinh 2r)}^{1/2}
}}
\ \ ,
\\
\noalign{\smallskip}
b &=& {\displaystyle{
{ {(2\sinh 2r)}^{1/2} p_r \over {2\cosh 2r} }
}}
\ \ ,
\end{array}
\end{equation}
where $r>0$ and $p_r$ is arbitrary. The symplectic structure
(\ref{omegac1}) becomes
\begin{equation}
\Omega = dp_r \wedge dr
\ \ .
\label{omega-across}
\end{equation}
Conjugating the holonomies (\ref{c1}) by the boost
$\left( \exp \left( \pm \casehalf \ln (\tanh r) K_1\right), 0
\right)$ yields
\begin{equation}
R_A = P
\ \ , \ \
w_A =
{p_r \over 2 \cosh 2r}
\left(
\begin{array}{c}
\pm e^{-r}
\\
0 \\
e^r
\end{array}
\right)
\ \ , \ \
R_B =
\exp( \pm e^{-r} K_0 + e^r K_2)
\ \ , \ \
w_B = 0
\ \ .
\label{kc1}
\end{equation}
Now, the holonomies (\ref{kc1}) continue smoothly to $r\le0$.
At $r=0$, (\ref{kc1})~coincides with~(\ref{c2}),
with the upper and lower signs matching,
provided we set $p=p_r$. For $r<0$, conjugating (\ref{kc1}) by
$\left( \exp \left( \mp \casehalf \ln (-\tanh r) K_1\right), 0
\right)$ yields the holonomies~(\ref{c4}), provided we set
\begin{equation}
\begin{array}{rcl}
{\tilde \lambda} &=&
{\displaystyle{
\pm {(-2\sinh 2r)}^{1/2}
}}
\ \ ,
\\
\noalign{\smallskip}
{\tilde b} &=&
{\displaystyle{
\mp { {(-2\sinh 2r)}^{1/2} p_r \over {2\cosh 2r} }
}}
\ \ .
\end{array}
\label{newpara2}
\end{equation}
Because ${\tilde \lambda}$ in (\ref{c4}) should be understood as an
angular parameter, equations (\ref{newpara2}) need to be interpreted
appropriately; for us it is sufficient to note that they have a
unambiguous meaning for $-\casehalf \arsinh (\pi^2/2) <r<0$,
separately for the upper and lower signs. The expression
(\ref{omega-across}) for the symplectic structure is valid for
$-\casehalf \arsinh (\pi^2/2) <r<\infty$ for both upper and lower
signs, and clearly coincides with (\ref{omega4c}) for $r<0$.
We have therefore
constructed a set of local coordinate systems from which it is seen
that the set $C_1 \cup C_2 \cup C_4$ is a smooth non-Hausdorff
manifold. It can be viewed as the cotangent bundle $T^*\bbase$ over a
tadpole-like non-Hausdorff base manifold~$\bbase$ (Figure~3).
$\bbase$~consists of the base space of $C_1$ (open half-line) and the
base space of $C_4$ (open interval) glued together in a non-Hausdorff
fashion by the two points that constitute the $\o$ projection
of~$C_2$.

One can now quantize the cotangent bundle $T^*\bbase$ in the
geometric quantization framework of
Refs.\cite{woodhouse,AAbook-geom}, again closely following the
treatment of the torus theory in Ref.\cite{louma}. In particular, the
subtleties arising from the non-Hausdorff property of $\bbase$ can be
handled as in Ref.\cite{louma}. The resulting quantum theory contains
operators that induce transitions between the
quantum theories built from $C_1$ and~$C_4$.

\section{Conclusions and discussion}
\label{sec:discussion}

In this paper we have investigated a connection formulation of 2+1
gravity that generalizes Witten's formulation\cite{witten1} to
nonorientable three-manifolds and to the full four-component 2+1
Poincare group $\io$. We first defined the theory, for a general
three-manifold~$M$, as a theory of flat connections in $\io$ bundles
over~$M$, and we discussed in some detail the notion of gauge
transformations and the recovery of spacetime metrics. We then
defined a class of bundles as orientation compatible iff the
(potential) nonorientability of $M$ intertwines with the (potential)
nontriviality of the bundle in a certain way. It was shown that for
orientation compatible bundles the theory has a natural action
principle, which reduces to that given in Ref.\cite{witten1} for
$\ioc$ bundles over oriented manifolds.

We next specialized to $M = \BbbR \times \KB$, where $\KB$ is the
Klein bottle. We analyzed in detail several of the connected
components of the solution space~$\miso$, including all the seven
components that arise from orientation compatible bundles. We
demonstrated that four of these seven components contain points from
which one can recover nondegenerate spacetime metrics on $\BbbR
\times \KB$. Some of these spacetimes are obtained by taking the
quotient of the 2+1 Minkowski space $\mink$ under the action of the
holonomies of the connection;
to obtain the others, one first removes
from $\mink$ a geodesic, passes to the universal covering space, and
then takes the quotient with respect to certain isometries that can
be associated with $\io$ elements only in a local sense. In
particular, from one of the connected components of
$\miso$ we recovered spacetime metrics in which
the induced metric on
$\KB$ is positive definite.

For the orientation compatible bundles, we used the Hamiltonian
decomposition of the action to define a
symplectic structure on the fields as in the orientable
case\cite{witten1,AAbook2}, and we then pulled this
symplectic structure back to a symplectic structure
on the associated components of~$\miso$.
This symplectic structure allows one to interpret
these components of~$\miso$, after excision of certain
singular subsets, as cotangent bundles over
the ``base spaces" where the $\io$
holonomies lie in $\o$. One can thus approach quantization of these
components of $\miso$ via the geometric quantization methods of
Refs.\cite{woodhouse,AAbook-geom}. For the component of $\miso$ that
was found to yield metrics with spacelike Klein bottles, the
resulting quantum theories are closely analogous to those that were
constructed for the $\ioc$ connection theory on $\BbbR \times T^2$ in
Ref.\cite{louma}. Extending the terminology of Ref.\cite{AAbook2},
one recovers two distinct ``spacelike sector" quantum theories and
one ``timelike sector" quantum theory. Further, there exists a
natural larger quantum theory that incorporates both the ``timelike
sector" theory and one of the ``spacelike sector" theories, and
contains operators that induce transitions between these
two smaller theories.

When defining the solution space~$\miso$, we chose to treat
both the large gauge transformations and the large diffeomorphisms as
symmetries rather than as gauge. Treating the large diffeomorphisms
as gauge would mean taking the quotient of $\miso$ with respect to
the action of the outer automorphisms of the fundamental group of the
Klein bottle; treating the large gauge transformations as gauge would
mean taking the quotient of $\miso$ with respect to conjugation by
$\io/\ioc$. Either option would result into discrete identifications
on~$\miso$. In the parametrizations of Section~\ref{sec:kbconnsol},
the identifications are straightforward to implement by restricting
the ranges of the parameters. For example, in $C_1$ (\ref{c1}) either
option would result into the restriction $b\ge0$. However, these
identifications are not compatible
with the interpretation of certain
components of $\miso$ (including~$C_1$) as cotangent bundles in the
manner of Section~\ref{sec:quantization}, since
the prospective momenta then no longer
take values in all of~$\BbbR$.
Handling these identifications in the quantum theory would
therefore require new input.

As the torus is a double cover of the Klein bottle (see
Appendix~\ref{app:kb}), there exists a natural map
${\cal P}\colon \miso \to \misotorus$,
where $\misotorus$ is the solution space of the
$\io$ connection theory on $\BbbR\times T^2$.
In terms of pairs of
$\io$ elements, in the notation of
Section~\ref{sec:kbconnsol}, this map is given by $(A,B)\mapsto
(A^2,B)$. The images of $\moo$, $\mpo$, and $\mto$ under
${\cal P}$ lie in the component
$\misotorusconn$ that comes from the $\ioc$
connection theory\cite{louma}. The symplectic
form on $\misotorusconn$ can therefore
be pulled back with ${\cal P}^*$ to two-forms on $\moo$, $\mpo$,
and~$\mto$. For $\mpo$ and~$\mto$, this yields
symplectic forms
that agree, up to an overall numerical factor, with the symplectic
forms obtained from our action~(\ref{action}).
For~$\moo$, on the other hand, the resulting two-form
is identically vanishing. We also see that
${\cal P}$ gives rise to a map that
takes all our nondegenerate metrics on
$\BbbR\times\KB$ to nondegenerate metrics on $\BbbR\times
T^2$. In terms of our local coordinates $(x,y)$, this map means that
the coordinates become identified according to $(x,y) \sim (x+1,y)
\sim (x,y+2)$.

In this paper we have not attempted to investigate directly
a metric formulation of Einstein gravity on $\BbbR \times \KB$.
If the induced metric on the Klein bottle is assumed to be
spacelike, a metric formulation could presumably
be analyzed by the methods of
Refs.\cite{moncrief1,mess,hosoya-nakao1}, and it appears likely that
all the classical solutions would be isometric to the
spacetimes~(\ref{c1fmetric1}), (\ref{c5fmetric1}),
and~(\ref{c3fmetric1})\cite{mess-priv}. If this is true, one could
eliminate the supermomentum constraints by adopting a spatially
locally homogeneous slicing, and one would then be led to the
``minisuperspace" metric theory for $\BbbR \times \KB$ discussed in
Refs.\cite{louru,giulo1}. For the nonstatic solutions
(\ref{c1fmetric1}) and~(\ref{c5fmetric1}), one could further solve
the remaining super-Hamiltonian constraint by adopting the York time
gauge, arriving at unconstrained ``square root Hamiltonian" theories
as in Ref.\cite{moncrief1}. The correspondence between metric
quantization and connection quantization could then be investigated
by methods that have been utilized for $\BbbR\times T^2$ in
Refs.\cite{carlip1,anderson,carlip2,carlip3}.

Our method to demonstrate the existence of nondegenerate spacetime
metrics was to explicitly construct such metrics from points in the
solution space of the connection theory. It would be interesting to
understand whether our collection of nondegenerate metrics is in any
sense an exhaustive one, and whether there is some easily
characterizable property of the connection theory that determines
which connected components of the solution space yield
nondegenerate metrics. For the $\ioc$ connection theory on
$\BbbR\times\Sigma$, where $\Sigma$ is a
closed orientable two-surface
of genus $g>1$, such a property is known: nondegenerate metrics with
spacelike $\Sigma$ are obtained precisely when the bundle has maximal
(or minimal) Euler class\cite{witten1,mess,amano}.

One is prompted to ask whether, for a general three-manifold~$M$,
orientation compatibility of the bundle might be a necessary
condition for recovering nondegenerate metrics from the connection
theory. An intuitive idea suggesting an affirmative answer is that
one might expect the spacetime to be always the quotient
of some domain in $\mink$ with respect to the action of the holonomy
group; this is known to be the case for
$\BbbR\times\Sigma$ with $\Sigma$ closed,
orientable, and spacelike\cite{mess}.
However, we have found the $\io$
connection theory on $\BbbR\times\KB$ to yield certain nondegenerate
metrics that come from quotienting the covering space of a multiply
connected subset of $\mink$ with respect to isometries that cannot be
globally interpreted as elements of $\io$. In Appendix
\ref{app:torustheory} we shall show that a similar observation holds
for a family of nondegenerate metrics arising from the $\io$ theory
on $\BbbR\times T^2$, such that the holonomy group is in
$\ioc\cup\iotp$ but not in $\ioc$. A~similar observation holds
already in the $\ioc$ theory on $\BbbR\times T^2$ when the tori are
not required to be spacelike\cite{louma}. What would be needed
is a better understanding of the relation
between the holonomy group of the connection and the isometry groups
that are employed in the quotient constructions.

In this paper we have given an action
principle only for orientation compatible bundles.
This meant that we were able to
introduce a symplectic structure only on those
components of the solution space that came from orientation
compatible bundles. It would be important to understand whether
this limitation could be removed. One possibility for examining
this issue might be to focus not on an action functional but
directly on the classical solution space. This avenue has been
developed in Refs.\cite{goldman1,karshon,huebschmann,jeffrey},
where a symplectic structure was
constructed on the spaces ${\rm Hom}(\Pi, G)/G$, where
$\Pi$ is a discrete group satisfying certain conditions, and
$G$ is a Lie group whose Lie algebra admits a symmetric
nondegenerate bilinear form that is invariant under the
adjoint action of~$G$. For $G=\ioc$, a
bilinear form satisfying the hypotheses
is~$P^I J_I$, in the notation of
Appendix~\ref{app:io}. When $\Pi= \pi_1(T^2) \simeq
\BbbZ\times\BbbZ$, it is straightforward to verify that the
approach of Refs.\cite{goldman1,karshon,huebschmann,jeffrey}
with $P^I J_I$ yields a symplectic structure that
agrees with the one obtained
from the action functional of the $\ioc$ theory
on $\BbbR\times T^2$\cite{carlip1,AAbook2,louma}.
This raises the hope that the construction of
Refs.\cite{goldman1,karshon,huebschmann,jeffrey}
could be generalized so as to apply
to the $\io$ theory on $\BbbR\times\KB$. One potential
difficulty here is, however, that the although $P^I
J_I$ is invariant under the adjoint action of $\ioc$ and
$\iotp$, it changes sign under the adjoint action of $\iop$ and
$\iot$.

In conclusion, we have seen that one connected component of the
$\io$ connection formulation of 2+1 gravity on
$\BbbR\times\KB$ is closely analogous to the single connected
component of the $\ioc$ connection formulation on $\BbbR\times
T^2$, both classically and quantum mechanically.
This connected component is arguably the
most interesting one from the viewpoint of spacetime metrics.
Although the classical $\io$ theory on $\BbbR\times\KB$
is in its own right solvable in explicit form,
it remains a subject to further work to fully examine
the possibilities for quantizing all the components of this
theory, as well as to fully assess the relevance of
all these components for spacetime metrics.

\acknowledgments

I~would like to thank Bruce Allen for asking the question which this
paper aims to answer, and for useful discussions. I~would also like
to thank John Friedman for helpful discussions and for his
constructive comments on the manuscript, and
Ric Ancel, John Barrett, Steve Carlip,
Domenico Giulini, Ted Jacobson, Jerzy Lewandowski, Eli Lubkin,
Don Marolf, Geoff Mess, and Alan Rendall for useful discussions and
correspondence.
This work was supported in part by the NSF grant
PHY91-05935.

\appendix
\section{$\io$}
\label{app:io}

In this appendix we collect some properties of $\io$ and establish
our notation.

The 2+1 dimensional Poincare group $\io$ can be defined as the group
of pairs $(R,w)$, where $R$ is an $\o$ matrix and $w$ is a column
vector with three real entries. The group multiplication law is
\begin{equation}
(R_2,w_2) \cdot (R_1,w_1) =
(R_2 R_1, R_2 w_1 + w_2)
\ \ .
\label{iomult}
\end{equation}
When points in the 2+1 dimensional Minkowski space $\mink$ are
represented by column vectors as $v={(T,X,Y)}^T$ and the entries are
the usual Minkowski coordinates associated with the line element
$ds^2 = - dT^2 + dX^2 + dY^2$, the action of a group element $(R,w)$
on $\mink$ is
\begin{equation}
(R,w) \colon \ \ v \mapsto Rv + w
\ \ .
\label{iomap}
\end{equation}

$\io$ is the semidirect product of the Lorentz subgroup $\o$, in
which the elements are of the form $(R,0)$, and the translational
subgroup, in which the elements are of the form $(\openone,w)$, where
$\openone$ stands for the $3\times3$ identity matrix.
$\io$ consists of the component of the identity $\ioc$, which in
(\ref{iomap}) preserves both space and time orientation, and the
disconnected components $\iop$, $\iot$, and~$\iotp$, which reverse
respectively the spatial orientation, time orientation,
and both space and time orientation. The corresponding four
components of $\o$ are denoted by $\oc$, $\op$, $\ot$, and $\otp$.

A basis for the Lie algebra of $\io$ is provided by the elements
$J_I$ and $P_I$ that can be respectively identified as the standard
bases of the Lorentz and translational subalgebras.
The uppercase Latin index takes values in $\left\{0,1,2\right\}$,
and the Lie brackets are
\begin{equation}
\begin{array}{rcl}
\left[ J_I , J_J \right] &=& \epsilon^K{}_{IJ} J_K
\\
\left[ J_I , P_J \right] &=& \epsilon^K{}_{IJ} P_K
\\
\left[ P_I , P_J \right] &=& 0
\ \ ,
\end{array}
\end{equation}
where $\epsilon^I{}_{JK}$ is obtained from the totally antisymmetric
symbol $\epsilon_{IJK}$ by raising the index with the 2+1 Minkowski
metric $\eta_{IJ}={\rm{diag}}(-1,1,1)$. Our convention is
$\epsilon_{012}=1$. The Lorentz indices are raised and lowered with
$\eta_{IJ}$ throughout the paper.

The adjoint representation of the Lie algebra of $\oc$ is spanned
by the matrices $K_J$ whose components are ${\left(K_J\right)}^I
{\vphantom{\left(K_I\right)}}_L=\epsilon^I{}_{JL}$.
Using the $\su$ parametrization of $\oc$\cite{bargmann,carmeli},
it is straightforward to verify that every
matrix in $\oc$ can be written as $\exp(v^I K_I)$,
where $v^I v_I \ge -\pi^2$.
The only redundancy in this parametrization is
that when $v^I v_I = -\pi^2$, $v^I$ and $-v^I$ give the same element.
For the use of Section \ref{sec:kbconnsol} we note the
parametrizations of a rotation, a boost and a null rotation
respectively as
\begin{equation}
\begin{array}{rcl}
&&\exp(v K_0) =
\left(
\begin{array}{ccc}
1 & 0 & 0 \\
0 & \cos v & -\sin v \\
0 & \sin v  & \cos v
\end{array}
\right)
\ \ , \ \
\exp(v K_2) =
\left(
\begin{array}{ccc}
\cosh v & \sinh v & 0 \\
\sinh v & \cosh v & 0 \\
0 & 0 & 1
\end{array}
\right)
\ \ ,
\\
\noalign{\bigskip}
&&\exp \left( v (\pm K_0+K_2) \right) =
\left(
\begin{array}{ccc}
1 + \half v^2 & v & \mp \half v^2 \\
v & 1 & \mp v \\
\pm \half v^2 & \pm v & 1 - \half v^2
\end{array}
\right)
\ \ .
\end{array}
\label{sopara}
\end{equation}

\section{Klein bottle}
\label{app:kb}

The Klein bottle $\KB$ can be constructed as the quotient manifold
${\BbbR}^2/H$, where $H$ is the group of diffeomorphisms of
${\BbbR}^2=\{(x,y)\}$ generated by the two elements
\begin{mathletters}
\label{abmap}
\begin{eqnarray}
{\bar a} \colon && (x,y) \mapsto (-x,y+1)
\ \ ,
\label{amap}
\\
{\bar b} \colon && (x,y) \mapsto (x+1,y)
\ \ .
\label{bmap}
\end{eqnarray}
\end{mathletters}%
As ${\bar{a}}$ reverses the orientation
of~${\BbbR}^2$, $\KB$ is nonorientable.
Mapping ${\BbbR}^2$ into a fundamental domain,
$\KB$~can be visualized as the closed square ${\overline
Q}=\left\{(x,y)\in
{\BbbR}^2 \mid 0\le x \le 1,\; 0\le y \le 1\right\}$, with the
vertical boundaries identified parallelly as $(0,y)\sim(1,y)$ and the
horizontal boundaries identified antiparallelly as
$(x,0)\sim(1-x,1)$.

The quotient construction implies
that ${\BbbR}^2$ is the universal covering space of~$\KB$,
and that the fundamental group $\pi_1(\KB):=\pi$
is isomorphic to~$H$. We denote
by $(a,b)$ a pair of generators of $\pi$ that corresponds to the
pair $({\bar a},{\bar b})$. From (\ref{abmap}) we have the relation
\begin{equation}
baba^{-1}=e
\ \ ,
\label{pirelation}
\end{equation}
where $e$ stands for the identity.
Conversely, $\pi$ can be defined as the discrete group generated by
two elements with the single
relation~(\ref{pirelation})\cite{massey}.

The elements of the automorphism group of~$\pi$, ${\rm Aut}(\pi)$,
are labeled by the triplets $(\epsilon,\eta,n)$,
where $\epsilon$ and $\eta$ take values in $\BbbZ_2$
and $n$ takes values in~$\BbbZ$.
The automorphisms act according to
\begin{equation}
(\epsilon,\eta,n) \colon
\ \
a\mapsto a^\epsilon b^n
\ \ , \ \
b\mapsto b^\eta
\ \ ,
\label{piauto}
\end{equation}
and the multiplication law is
$(\epsilon',\eta',n')\cdot(\epsilon,\eta,n)
= (\epsilon\epsilon',\eta\eta',n'+\eta'n)$.
We therefore have ${\rm Aut}(\pi) \simeq \BbbZ_2 \times \left(
\BbbZ_2 \times_{\rm S} \BbbZ \right)$, where $\times_{\rm S}$ stands
for a semidirect product. The subgroup of inner automorphisms, ${\rm
Inn}(\pi)$, consists of the elements for which $\epsilon=1$ and $n$
is even. The quotient group ${\rm Out}(\pi):={\rm Aut}(\pi)/{\rm
Inn}(\pi)$ is thus isomorphic to $\BbbZ_2 \times \BbbZ_2$, and the
homomorphism from ${\rm Aut}(\pi)$ to ${\rm Out}(\pi)$ is
$(\epsilon,\eta,n) \mapsto (\epsilon, e^{i\pi n}) \in \BbbZ_2 \times
\BbbZ_2$.

Taking the quotient of ${\BbbR}^2$ with respect to the discrete group
generated by the two diffeomorphisms ${\bar{a}}^2$ and ${\bar{b}}$
gives the torus~$T^2$. It follows that $T^2$ is a double cover
of~$\KB$.

\section{Lorentz subspace of~$\mpo$}
\label{app:mmpo}

In this appendix we consider the subspace $m\subset\mpo$~(\ref{mpo})
in which the holonomies are in $\o$. We have
\begin{equation}
m = {\tilde m}/\oc
\ \ ,
\label{mmpo1}
\end{equation}
where
\begin{equation}
{\tilde m} = \{ \, (R_A,R_B)\in \op\times\oc \mid R_B R_A R_B
R_A^{-1}=\openone \, \}
\ \ .
\label{mmpo2}
\end{equation}
We shall give for each point in $m$ a unique representative in
${\tilde m}$. The extension of this analysis to $\mpo$ is
straightforward.

We write $R_A = PF$, where $F\in\oc$ and $P={\rm diag}(1,-1,1)$. We
parametrize $R_B$ as $R_B=\exp(v^IK_I)$, where $v^I$ is interpreted
as a Lorentz-vector. The relation
$R_B R_A R_B R_A^{-1}=\openone$ then takes the form
\begin{equation}
\label{rel1}
\exp[{(Fv)}^IK_I] = \exp[{(Pv)}^IK_I]
\ \ .
\end{equation}

Suppose first that $v$ is spacelike, $v^Iv_I>0$. By $\oc$ conjugation
we can uniquely set $v={(0,0,\lambda)}^T$, where $\lambda>0$.
Then $Pv=v$, and (\ref{rel1}) implies $Fv=v$.
This means that $F$ is either the
identity or a boost that fixes~$v$.
If $R$ is a boost that fixes~$v$,
conjugation by $R$ leaves $R_B$ invariant but sends
$PF$ to $RPFR^{-1}= P(PRP)FR^{-1}=
PR^{-1}FR^{-1}= PFR^{-2}$. Choosing $R$ suitably one can thus set
$F=\openone$.

Suppose next that $v$ is nonzero null, $v^Iv_I=0$ but $v\neq0$. The
situation is analogous to the previous one. By $\oc$ conjugation we
can set $v={(\pm1,0,1)}^T$, after which (\ref{rel1}) implies $Fv=v$.
$F$~can then be conjugated to $\openone$ by a null rotation that
fixes~$v$.

Suppose next that $v$ is timelike, $v^Iv_I<0$. The situation is
analogous to those above. By $\oc$ conjugation we can uniquely set
$v={({\tilde\lambda},0,0)}^T$, where $-\pi<{\tilde\lambda}<0$ or
$0<{\tilde\lambda}\le\pi$. Then $Pv=v$, (\ref{rel1})~implies $Fv=v$,
and $F$ can be conjugated to $\openone$ by a rotation that fixes~$v$.

What remains is the case $v=0$. The relation (\ref{rel1}) is now an
identity. Conjugation by
$R=\exp(-u^IK_I)\in\oc$ sends $PF$ to
$RPFR^{-1}= P(PRP)FR^{-1}
= P \exp[{(Pu)}^IK_I] F \exp(u^IK_I)$.
This conjugation is perhaps most easily analyzed in the $\su$
parametrization of $\oc$\cite{bargmann,carmeli}, building $R$ from
infinitesimal conjugations by integration. One finds that $F$ can be
uniquely conjugated to $\exp(\mu K_1)$ with $\mu\geq0$.

\section{$\io$ connection theory on $\BbbR\times T^2$}
\label{app:torustheory}

In this appendix we discuss briefly the connection theory with the
full gauge group $\io$ on the manifold $\BbbR\times T^2$.

As $\pi_1(T^2) \simeq \BbbZ \times \BbbZ$, the classical solution
space is parametrized by pairs of commuting $\io$ elements modulo
overall $\ioc$ conjugation. The component where all the holonomies
are in $\ioc$ reduces to the theory considered in Ref.\cite{louma}.
Here we wish to investigate the components where the holonomies are
in $\iotp$ and $\ioc$. It is sufficient to consider the space
\begin{equation}
\misotorusone =
\{ \, (A,B)\in \ioc\times\iotp \mid ABA^{-1}B^{-1}=E \, \}
/ \ioc
\ \ .
\label{misotorusone}
\end{equation}
The space $\misotorustwo$, where $(A,B)\in
\iotp\times\iotp$, is isomorphic to $\misotorusone$ via a large
diffeomorphism of the torus.

With the help of the results in Appendix~\ref{app:mmpo} about the
$\oc$ conjugacy classes in $\op$, analyzing $\misotorusone$ is
straightforward. $\misotorusone$ consists of two connected
components, which we denote by $T_1$ and~$T_2$.
Representatives for the points in $T_1$ are given by
\begin{equation}
R_A =
\exp(\mu K_1)
\ \ , \ \
w_A =
\left(
\begin{array}{c}
0 \\
b \\
0
\end{array}
\right)
\ \ , \ \
R_B =
- P \exp(\lambda K_1)
\ \ , \ \
w_B =
\left(
\begin{array}{c}
0 \\
a \\
0
\end{array}
\right)
\ \ ,
\label{t1}
\end{equation}
and the parametrization becomes unique after the identification
$(\lambda,\mu,a,b)\sim(-\lambda,-\mu,-a,-b)$. Representatives for the
points in $T_2$ are given by
\begin{equation}
R_A =
\exp(\pi K_0)
\ \ , \ \
w_A =
\left(
\begin{array}{c}
0 \\
b \\
0
\end{array}
\right)
\ \ , \ \
R_B =
- P \exp(\lambda K_1)
\ \ , \ \
w_B = 0
\ \ ,
\label{t2}
\end{equation}
and the parametrization becomes unique after the identification
$(\lambda,b)\sim(-\lambda,-b)$.
The action (\ref{action}) endows $T_1$ and $T_2$
with the symplectic structures
\begin{mathletters}
\label{omegat}
\begin{eqnarray}
T_1:&&\ \
\Omega =  db \wedge d\lambda - da \wedge d\mu
\ \ ;
\label{omegat1}
\\
T_2:&&\ \
\Omega = db \wedge d\lambda
\ \ .
\end{eqnarray}
\end{mathletters}%
$T_1$~and $T_2$ can therefore each be regarded as a
cotangent bundle, if the points with $\lambda=\mu=0$
and $\lambda=0$ are respectively excised. Methods of geometric
quantization\cite{woodhouse,AAbook-geom} can thus be applied
as in Ref.\cite{louma}.

We now concentrate on~$T_1$. We envisage the torus as the closed
square ${\overline Q}=\left\{(u,v)\in {\BbbR}^2 \mid 0\le u \le 1,\;
0\le v \le 1\right\}$, with the horizontal boundaries identified as
$(u,0)\sim(u,1)$ and the vertical boundaries identified as
$(0,v)\sim(1,v)$. In the pairs $(A,B)\in\io\times\io$ used
in~(\ref{misotorusone}), we identify the
first member as coming from a closed loop at constant $u$ and the
second member as coming from a closed loop at constant~$v$. Consider
now a bundle that admits a local chart
$(\BbbR \times Q)\times\io$ such that
the transition function across the horizontal
boundaries $v=0$ and $v=1$ is the identity but the transition
function across the vertical boundaries $u=0$ and $u=1$ is
$(-P,0)$. In this chart, consider the connection
one-form
\begin{equation}
A^1 = \lambda du + \mu dv
\ \ , \ \
e^1 = a du + b dv
\ \ .
\label{t1c}
\end{equation}
(\ref{t1c})~clearly defines on the bundle a connection whose
holonomies are~(\ref{t1}).

To obtain a nondegenerate triad, we change the chart by the
transition function
$(\openone, {(t \cos(\pi u),-t\sin(\pi u),0)}^T)$. The new triad is
\begin{equation}
\begin{array}{rcl}
e^0 &=& \cos(\pi u) dt
+ t \sin(\pi u) [ (\lambda-\pi) du + \mu dv ]
\ \ ,
\\
e^1 &=& a du + b dv
\ \ ,
\\
e^2 &=& -\sin(\pi u) dt
- t \cos(\pi u) [ (\lambda+\pi) du + \mu dv ]
\ \ ,
\end{array}
\label{t1ftriad1}
\end{equation}
and in the new chart the transition functions across the
boundaries are the same as in the old chart. Squaring the
triad (\ref{t1ftriad1}) gives the metric
\begin{eqnarray}
ds^2 &=&
- \cos(2\pi u) dt^2
+ 2\pi \sin(2\pi u) t dt dv
+ {(a du + b dv)}^2
\nonumber
\\
&&+ t^2 \left\{ \cos(2\pi u)
\left[ {(\lambda du + \mu dv)}^2 + \pi^2 du^2 \right]
+ 2\pi (\lambda du + \mu dv) du \right\}
\ \ .
\label{t1fmetric1}
\end{eqnarray}
For $|b\lambda - a\mu| < \pi|b|$, the metric (\ref{t1fmetric1}) is
nondegenerate for $t>0$, and it clearly defines a nondegenerate
metric on $\BbbR \times T^2$. The local coordinate transformation
\begin{equation}
\begin{array}{rcl}
U &=& 2^{1/2} t
\exp(\lambda u + \mu v) \sin(\pi u + \pi/4)
\ \ ,
\\
V &=& 2^{1/2} t
\exp(-\lambda u - \mu v) \cos(\pi u + \pi/4)
\ \ ,
\\
Y &=& a u + b v
\ \ ,
\end{array}
\label{tnulltrans}
\end{equation}
brings the metric (\ref{t1fmetric1}) to the explicitly flat double
null form~(\ref{minknullmetric}). It is then seen as in subsection
\ref{subsec:mTbundle} that the spacetime (\ref{t1fmetric1}) is
obtained by cutting a spacelike geodesic from Minkowski space, going
to the universal covering space, and taking the quotient with respect
to two isometries that can be locally represented by the two
holonomies~(\ref{t1}).

\begin{figure}
\caption{The projection of the component
$\bigcup_{i=1}^6 A_i \subset \moo$ into
$\hbox{Hom}(\pi,\oc)/\oc$.
We denote the projections of $A_i$ respectively by
$A'_i$. $A'_4\simeq\BbbR\cup\BbbR$ is completed into a smooth circle
by the two points $A'_3$ and~$A'_5$, at which also the two lines
$A'_1$ and $A'_6$ join the circle. The two points that constitute
$A'_2$ are close to $A'_1$ and $A'_3$ in a non-Hausdorff way.}
\label{fig:moo}
\end{figure}

\begin{figure}
\caption{The projection of $\mpo$ into $\hbox{Hom}(\pi,\o)/\oc$.
The projections of the sets $C_i$ are denoted
respectively by~$C'_i$. $C'_1$, $C'_4$, and $C'_5$ all meet at the
point $C'_3$.}
\label{fig:mpo}
\end{figure}

\begin{figure}
\caption{The tadpole-like non-Hausdorff manifold~$\bbase$, consisting
of the sets~$C'_1$, $C'_2$, and $C'_4$ of
Figure~\protect\ref{fig:mpo}. One end of the open half-line $C'_1$ is
glued to both ends of the open interval $C'_1$ by the two points that
constitute~$C'_2$. $\bbase$~is a manifold, but the two points
constituting $C'_2$ do not have disjoint neighborhoods.}
\label{fig:tadpole}
\end{figure}

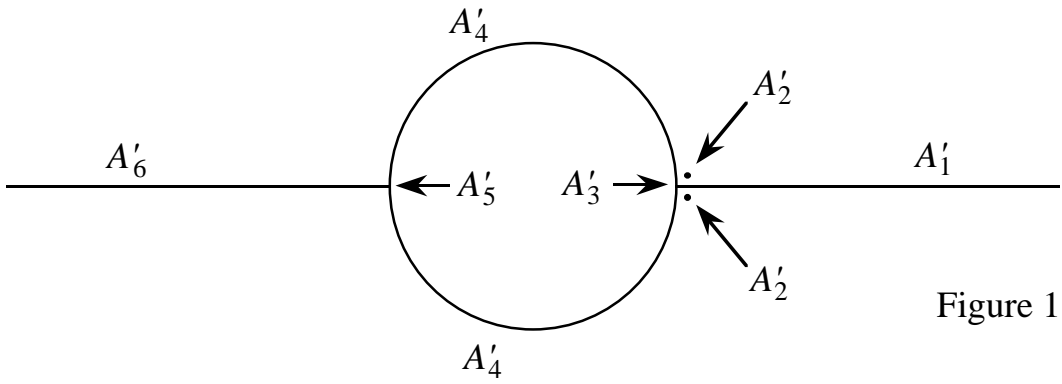
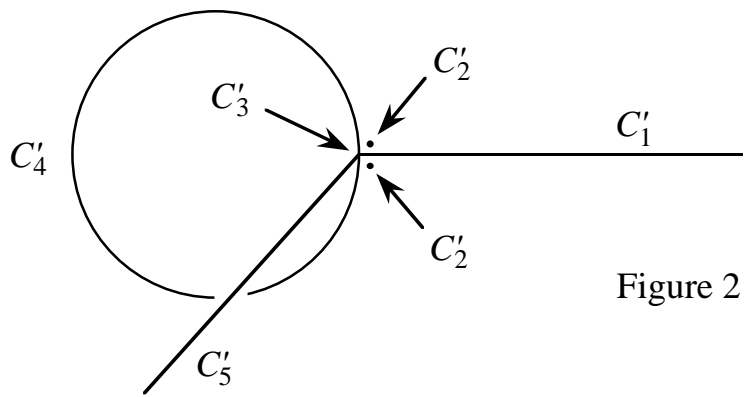
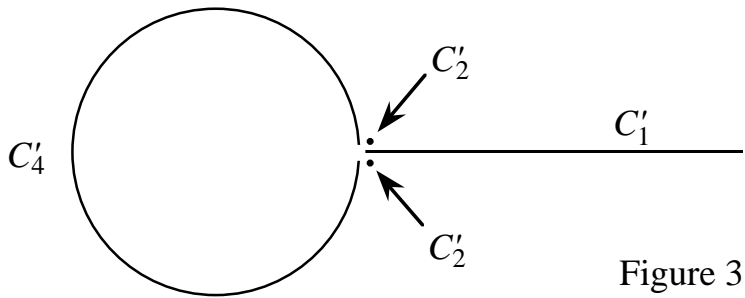
\end{document}